\documentclass[aps,prb,reprint,superscriptaddress]{revtex4-2}

\usepackage{amsmath, amssymb}
\usepackage{bm}
\usepackage{graphicx}
\usepackage{color}
\usepackage{dcolumn}
\usepackage{bm}
\usepackage[hidelinks]{hyperref}
\hypersetup{
  colorlinks   = true, 
  urlcolor     = blue, 
  linkcolor    = blue, 
  citecolor   = red 
}
\usepackage{multirow}
\usepackage{comment}

\usepackage{soul}

\usepackage{threeparttable}

\allowdisplaybreaks[1]

\begin{document}

\title{
Coupled-cluster approach to vibronic effects in resonant inelastic x-ray scattering of quantum materials: Application to a $5d^1$ rhenium oxide
}

\author{Teruki Matsuzaki}
\affiliation{Department of Chemistry, KU Leuven, Celestijnenlaan 200F, B-3001 Leuven, Belgium}
\affiliation{Graduate School of Science and Engineering, Chiba University, 1-33 Yayoi-cho, Inage-ku, Chiba-shi, Chiba 263-8522, Japan} 

\author{Liviu F. Chibotaru}
\email[]{liviu.chibotaru@kuleuven.be}
\affiliation{Theory of Nanomaterials Group, KU Leuven, Celestijnenlaan 200F, B-3001 Leuven, Belgium}

\author{Maristella Alessio}
\email[]{maristella.alessio@kuleuven.be}
\affiliation{Department of Chemistry, KU Leuven, Celestijnenlaan 200F, B-3001 Leuven, Belgium}

\author{Naoya Iwahara}
\email[]{naoya.iwahara@gmail.com}
\affiliation{Graduate School of Engineering, Chiba University, 1-33 Yayoi-cho, Inage-ku, Chiba-shi, Chiba 263-8522, Japan}

\begin{abstract}
First-principles analysis of the spectroscopic signatures of correlated quantum materials poses significant challenges due to the interplay between spin-orbit and vibronic couplings, as well as the need to describe both dynamic and static electron correlation to reach decent accuracy. 
In this work, we apply the equation-of-motion coupled-cluster (EOM-CC) method to derive the spin-orbit-lattice entangled vibronic states and predict the Re $L_3$ edge resonant inelastic x-ray scattering (RIXS) spectra of Ba$_2$MgReO$_6$.
The EOM-CC yields interaction parameters in close agreement with those extracted from RIXS spectra, with errors of less than 5\%.
In particular, the EOM-CC method allowed us to determine the weak vibronic coupling to the $T_{2g}$ vibrations, which is difficult to address experimentally.
The simulated spectra indicate that vibronic coupling to the $T_{2g}$ modes gives rise to a shoulder on the elastic peak. 
Going beyond the conventional treatment, which focuses solely on $E_g$ modes, we show that vibronic couplings to both $T_{2g}$ and $E_g$ modes are required to account for the fine structure of the RIXS spectra.
This work demonstrates that the EOM-CC method is a powerful tool for accurately predicting the complex local states at metal sites and spectroscopic signatures of correlated insulating materials. 
\end{abstract}

\maketitle

\section{Introduction}
\label{Sec_intro}
Heavy transition metal compounds host unconventional magnetism \cite{Witczak-Krempa2014, Rau2016, Hermanns2018, Takagi2019, Motome2020Majorana, Takayama2021, Trebst2022, Chen2024, Pourovskii2025}.
Over the last two decades, the iridates and ruthenates have been the playground for exploring the Kitaev spin-liquid phase and Majorana fermions.
The $5d^1$ and $5d^2$ transition metal compounds have attracted recent attention as a new family of multipolar ordered materials. 
Unlike conventional multipolar phases observed in strongly spin-orbit coupled lanthanide and actinide compounds \cite{Santini2009, Onimaru2016, Suzuki2018, Rau2019}, in heavy transition metal compounds, the electron-phonon (vibronic) coupling affects multipolar phases. 

\begin{figure}[tb]
 \includegraphics[width=\linewidth, bb=0 0 926 422]{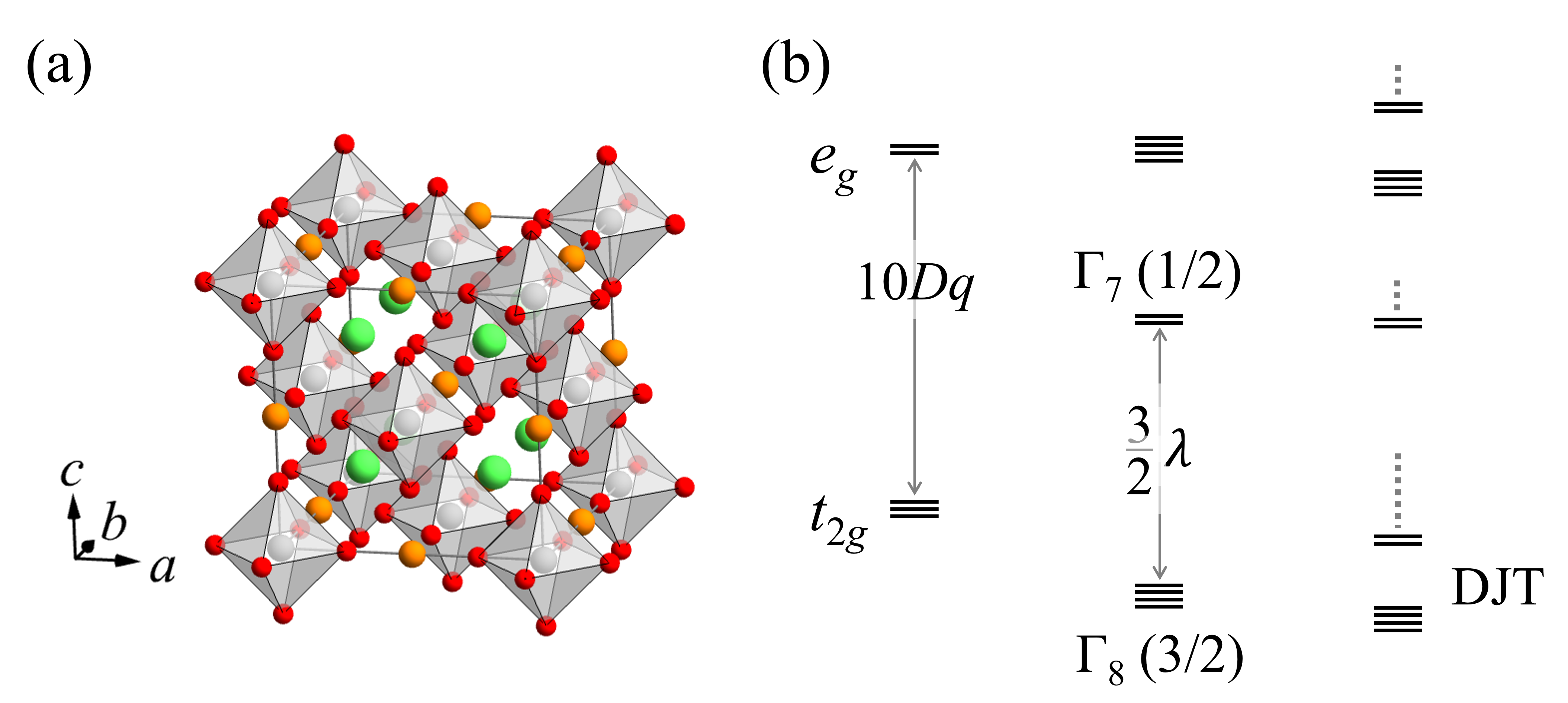}
 \caption{
 The conventional unit cell of Ba$_2$MgReO$_6$ and the energy spectra on a $5d^1$ site. 
 (a) The green, orange, light gray, and red spheres are Ba, Mg, Re, and O atoms, respectively. $a, b, c$ indicate the crystal axes of the conventional cell. 
 (b) In the descending order of energy, the ligand-field, spin-orbit coupling, and vibronic coupling (DJT effect) determine the nature of the local quantum states. 
 $10Dq$ and $\lambda$ are the ligand-field and spin-orbit coupling parameters. 
 $\Gamma_7$ and $\Gamma_8$ are the irreducible representations of the spin-orbit multiplet $j_\text{eff}=1/2$ and $j_\text{eff}=3/2$ states, respectively, in the $O_h$ group.  
 The dots in the last column indicate the presence of a large number of vibronic levels. 
 }
 \label{Fig_DP}
\end{figure}

The family of cubic $5d^1$ double perovskites exhibits various multipolar ordered phases induced by the spin-orbit and the vibronic couplings. 
In these compounds, the magnetic $5d^1$ metal ions in an octahedral environment form the face-centered cubic structure [Fig. \ref{Fig_DP}(a)].
On each $5d^1$ center, the ligand-field and spin-orbit coupling stabilize the $j_\text{eff} = 3/2$ multiplet
[Fig. \ref{Fig_DP}(b)], causing the emergence of the multipolar nature of the ordered phases.
The theories based on the spin-orbit and spin-orbital exchange interactions describe canted antiferromagnetic and antiferro-quadrupolar phase \cite{Chen2010, Svoboda2021, Kubo2023}, which appears in Ba$_2$NaOsO$_6$ and Ba$_2$MgReO$_6$ at low temperature \cite{Marjerrison2016, Lu2017, Liu2018, Willa2019, Hirai2019}, while fail to describe the high-temperature quadrupolar phase in Ba$_2$MgReO$_6$ \cite{Hirai2020, Soh2024, Muroi2025} and the ordered phase of Cs$_2$TaCl$_6$ \cite{Ishikawa2019, MansouriTehrani2023}. 
The deviation could arise due to vibronic effects as suggested by the structural transitions observed in Ba$_2$MgReO$_6$ \cite{Hirai2020, Soh2024, Muroi2025} and Cs$_2$TaCl$_6$ \cite{Ishikawa2019, MansouriTehrani2023}.
Density functional theory (DFT) studies also demonstrate that structural deformations are crucial for reproducing the canting angle of the magnetic moments \cite{FioreMosca2021, MansouriTehrani2021}.

The effect of vibronic coupling on the $5d^1$ site is not limited to inducing structural deformations, but can also give rise to spin-orbit-lattice entanglement. 
The $j_\text{eff}=3/2$ spin-orbit multiplet states couple to the Jahn-Teller (JT) active vibrations (Fig. \ref{Fig_JTmodes}) of the surrounding atoms, leading to the formation of their quantum entanglement, which is the dynamic Jahn-Teller (DJT) effect \cite{Englman1972, Moffitt1957, Iwahara2017, Iwahara2018, Iwahara2024}.
Due to the large distance between the $5d^1$ centers in the double perovskites and the resulting weak intersite interactions, the DJT effect on the metal sites persists. 
In DJT systems, the energy-level distribution differs significantly from the equally spaced electronic-vibrational type \cite{Longuet-Higgins1958, Muramatsu1978, Englman1972,  Iwahara2024}. 
Furthermore, the DJT effect affects the nature of the multipolar moments and their orderings, which qualitatively explain the puzzling ordered phases \cite{Iwahara2023}.

Recent resonant inelastic x-ray scattering (RIXS) measurements unveiled the presence of the DJT effect in the $5d^1$ double perovskites. 
The RIXS spectroscopy is able to capture various elementary excitations \cite{Ament2011, Gilmore2023, Mitrano2024, deGroot2024}, providing fundamental information on the complex energy levels of $5d$ transition metal compounds \cite{Taylor2017, Jungho2020, Warzanowski2023, Magnaterra2024, Okamoto2025, Frontini2025}.
The RIXS spectra of the $5d^1$ double perovskites show an unexpected asymmetry of the spin-orbit $j_\text{eff}=1/2$ peak and the presence of the shoulder structures near the elastic peak \cite{Frontini2024, Agrestini2024, Zivkovic2024, Iwahara2025}.
Theoretical analysis of the RIXS spectra of rhenium compounds demonstrated that these structures are the fingerprints of the JT dynamics. 
The asymmetric peak for the excited spin-orbit $j_\text{eff} = 1/2$ multiplet was detected in Re $L_3$ edge RIXS spectra \cite{Frontini2024}, accompanied by the low-energy peaks for the transitions between the vibronic levels in oxygen $K$-edge RIXS spectra \cite{Iwahara2025}.

However, the previous theoretical analysis of the RIXS spectra could not unravel the origin of the shoulder to the elastic peak in the Re $L_3$ edge RIXS spectra of Ba$_2$MgReO$_6$ \cite{Frontini2024}. 
Whether the spin-orbit-lattice entangled vibronic states are sufficient to describe the $5d^1$ sites or whether other factors play a crucial role should still be clarified.

The failure in interpreting the shoulder in the Re $L_3$ edge RIXS spectra could stem from the insufficient knowledge of the DJT effect of the embedded $5d^1$ ion sites. 
The $j_\text{eff} = 3/2$ multiplet states couple to the two types of JT-active modes [Fig. \ref{Fig_JTmodes}], while the employed model for the analysis included only the vibronic coupling to the $E_g$ modes \cite{Frontini2024, Iwahara2025}. 
This approximation was motivated by the general tendency of weaker vibronic coupling of the $T_{2g}$ modes to the $d$ orbitals compared to the $E_g$ modes and the difficulty in determining the weak vibronic coupling from the RIXS spectra.
To unravel the true vibronic nature of the $5d^1$ sites, we must revisit the analysis of the RIXS spectra using a reliable DJT model that incorporates all relevant vibronic couplings.

To determine the DJT model, first-principles calculations are indispensable.
Several groups, including us, have addressed the first-principles derivations of DJT models for embedded $5d^1$ ions using various quantum chemistry methods such as complete active space self-consistent field (CASSCF) with multireference configuration interaction (MRCI) \cite{Xu2016, Zivkovic2024, Soh2024}, CASSCF with a M{\o}ller-Plesset perturbation theory \cite{Iwahara2018}, and DFT with Hubbard I approximation \cite{FioreMosca2024}. 
All first-principles calculations undertaken so far demonstrate that the vibronic coupling to the $E_g$ modes is dominant, however, their analysis of the role of $T_{2g}$ modes is not quantitatively conclusive.

In this work, we theoretically analyze the Re $L_3$ edge RIXS spectra of Ba$_2$MgReO$_6$.
We obtain the dynamic Jahn-Teller model for $5d^1$ Re cluster with the state-of-the-art equation-of-motion coupled-cluster (EOM-CC) method.
The EOM-CC method provides a robust and efficient framework for treating open-shell systems, describing the multiconfigurational character of their wavefunctions within a single-reference formalism \cite{Krylov2008, Sneskov2012, Bartlett2012}. 
The EOM-CC approach provides a balanced treatment of all relevant target states, capturing both dynamical and non-dynamical correlation within a single computational step.
These methods have been extensively benchmarked for a wide range of transition-metal complexes and have demonstrated high accuracy in predicting state energies and magnetic properties
\cite{Orms2018, Pokhilko2019, Maristella2021, Maristella2023, Kahler2023, Eversdijk2025}.
Using the derived DJT model, we simulated the vibronic states and the Re $L_3$ RIXS spectra.
The calculated ligand-field, spin-orbit coupling, and vibronic coupling to the $E_g$ mode agree well with the estimated values from the Re $L_3$ RIXS spectra \cite{Frontini2024}. 
In addition, we reveal that vibronic coupling to the $T_{2g}$ mode gives rise to a shoulder of the elastic peak of the RIXS spectra. 
These results illustrate the effectiveness of the EOM-CC method for exploring the nature and spectroscopic signatures of quantum materials.

\section{Quantum states of an $5d^1$ octahedron}
\label{Sec_theory}
\subsection{Vibronic model Hamiltonian}
We begin by reviewing the vibronic interaction in embedded $5d^{1}$ ions in cubic double perovskites.
On the $5d^1$ sites, ligand-field, spin-orbit coupling, and vibronic coupling coexist. 
We describe these interactions in the descending order of the energy scale.

\begin{figure}[tb]
\includegraphics[width=0.8\linewidth, bb=0 0 510 275]{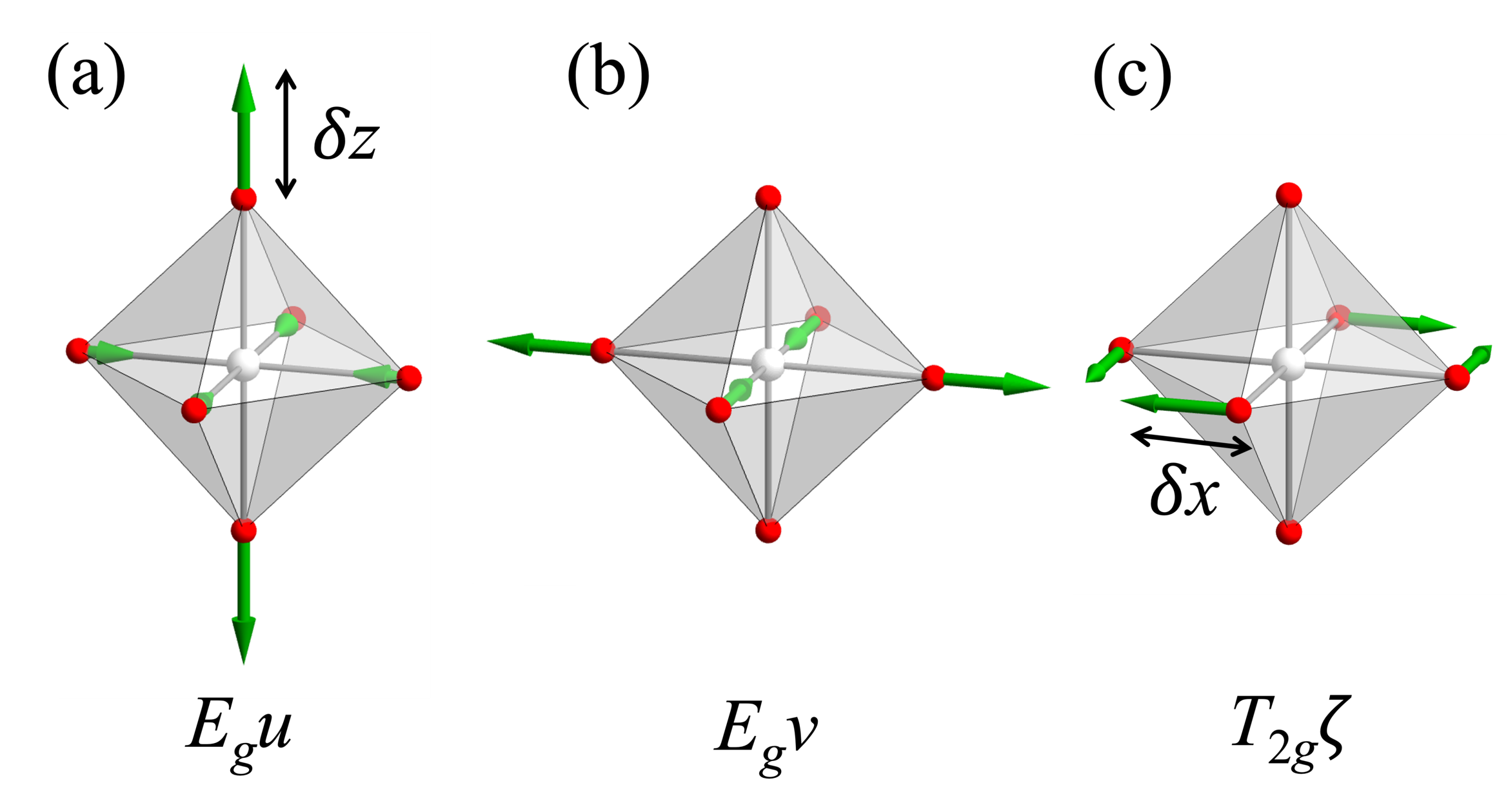}
 \caption{The JT active modes. (a) The $E_gu$ $(=2z^2-x^2-y^2)$ type, (b) $E_gv$ $(=x^2-y^2)$ type, and (c) one of the $T_{2g}$ modes ($\zeta = xy$ type), where the $x, y, z$ directions correspond to the positive $a, b, c$ directions in Fig. \ref{Fig_DP}, respectively. 
 Cyclic permutations of $xyz$ obtain the other two $T_{2g}$ modes.
 The green arrows indicate the displacement for the normal mode. 
 These displacements correspond to the positive direction of the normal coordinates. 
 }
\label{Fig_JTmodes}
\end{figure}

Our electronic Hamiltonian for an embedded $5d^1$ metal ion consists of the ligand-field and spin-orbit coupling:
\begin{align}
 \hat{H}_\text{el} &= \hat{H}_\text{LF} + \hat{H}_\text{SO}.
 \label{Eq_Hel}
\end{align}
The ligand-field Hamiltonian splits the $5d$ orbital levels into the $e_g$ and $t_{2g}$ orbital levels \cite{Sugano1970}:
\begin{align}
\hat{H}_\text{LF} &= 
10Dq 
\left(
\frac{3}{5}
 \sum_{\gamma=u,v} |\gamma\rangle \langle \gamma|
-
\frac{2}{5} \sum_{\gamma=\xi,\eta,\zeta} |\gamma\rangle \langle \gamma|
\right).
 \label{Eq_hLF}
\end{align}
Here, $10Dq$ ($>0$) is the ligand-field splitting parameter, and $|\gamma\rangle$ are the $5d$ orbital states. 
$\gamma$ $=u, v$ indicate the basis of $e_g$ irreducible representation, and $\gamma= \xi, \eta, \zeta$ the basis of $t_{2g}$ representation.
$\gamma = u, v, \xi, \eta, \zeta$ transform as $(2z^2-x^2-y^2)/\sqrt{6}$, $(x^2-y^2)/\sqrt{2}$, $\sqrt{2} yz$, $\sqrt{2} zx$, and $\sqrt{2} xy$, respectively, under the symmetry operations of the $O_h$ group. 
The octahedral ligand field makes the $t_{2g}$ orbital levels lower than the $e_g$ orbital levels. 

$\hat{H}_\text{SO}$ is the atomic spin-orbit coupling between the $5d$ orbital and spin angular momenta:
\begin{align}%
 \hat{H}_\text{SO} &= \lambda \sum_{\gamma\sigma} \sum_{\gamma'\sigma'} \sum_{\alpha = x, y, z} (\bm{l}_\alpha)_{\gamma\gamma'} \cdot (\bm{s}_\alpha)_{\sigma\sigma'} 
 |\gamma\sigma\rangle
 \langle\gamma'\sigma'|.
 \label{Eq_HSO}
\end{align}
Here, $\lambda$ ($>0$) is the spin-orbit coupling parameter, $\bm{l}_\alpha$ ($\alpha = x, y, z$) are the orbital angular momentum matrices,
\begin{align}
 \bm{l}_x &= 
 \begin{pmatrix}
  0 & 0 & \sqrt{3}i & 0 & 0 \\
  0 & 0 & i & 0 & 0 \\
  -\sqrt{3} i & -i & 0 & 0 & 0 \\
  0 & 0 & 0 & 0 & i \\
  0 & 0 & 0 & -i & 0 \\
 \end{pmatrix},
 \nonumber\\
 \bm{l}_y &= 
 \begin{pmatrix}
  0 & 0 & 0 & -\sqrt{3} i & 0 \\
  0 & 0 & 0 & i & 0 \\
  0 & 0 & 0 & 0 & -i \\
  \sqrt{3} i & -i & 0 & 0 & 0 \\ 
  0 & 0 & i & 0 & 0 \\
 \end{pmatrix},
 \nonumber\\
 \bm{l}_z &= 
 \begin{pmatrix}
  0 & 0 & 0 & 0 & 0 \\
  0 & 0 & 0 & 0 & -2i \\
  0 & 0 & 0 & i & 0 \\
  0 & 0 & -i & 0 & 0 \\
  0 & 2i & 0 & 0 & 0 \\
 \end{pmatrix},
\end{align}
with the basis orbitals $|u\rangle$, $|v\rangle$, $|\xi\rangle$, $|\eta\rangle$, $|\zeta\rangle$,
$\bm{s}_\alpha$ are the $s=1/2$ spin angular momentum matrices,
and $|\gamma\sigma\rangle$ are the $5d$ spin-orbital states. 

The $d$ orbitals couple to the JT active $E_g$ and $T_{2g}$ modes (Fig. \ref{Fig_JTmodes}) according to the selection rule.
The DJT Hamiltonian consists of the harmonic oscillator Hamiltonian for the JT active modes and the vibronic coupling:
\begin{widetext}
\begin{align}
 \hat{H}_\text{DJT} &= 
\sum_{\Gamma = E_g, T_{2g}} \sum_{\gamma \in \Gamma} \frac{1}{2}\left(\hat{P}_\gamma^2 + \omega_\Gamma^2 \hat{Q}_\gamma^2 \right)
+
 \left(
  |u     \rangle,
  |v     \rangle,
  |\xi   \rangle,
  |\eta  \rangle, 
  |\zeta \rangle 
  \right)
  \nonumber\\
  &\times
  \left[
  V_E
 \begin{pmatrix}
 -\hat{Q}_u   & \hat{Q}_v & 0 & 0 & 0 \\
  \hat{Q}_v & \hat{Q}_u   & 0 & 0 & 0 \\
  0 & 0 & -\frac{1}{2} \hat{Q}_u + \frac{\sqrt{3}}{2} \hat{Q}_v & 0 & 0 \\
  0 & 0 & 0 & -\frac{1}{2} \hat{Q}_u - \frac{\sqrt{3}}{2} \hat{Q}_v & 0 \\
  0 & 0 & 0 & 0 & \hat{Q}_u  \\
 \end{pmatrix}
 + 
 V_{T_2}
 \begin{pmatrix}
   0 & 0 & -\frac{1}{2} \hat{Q}_\xi & -\frac{1}{2} \hat{Q}_\eta & \hat{Q}_\zeta \\
   0 & 0 & \frac{\sqrt{3}}{2} \hat{Q}_\xi & -\frac{\sqrt{3}}{2} \hat{Q}_\eta & 0 \\
   -\frac{1}{2} \hat{Q}_\xi  &  \frac{\sqrt{3}}{2} \hat{Q}_\xi  & 0 & -\frac{\sqrt{3}}{2} \hat{Q}_\zeta & -\frac{\sqrt{3}}{2} \hat{Q}_\eta \\ 
   -\frac{1}{2} \hat{Q}_\eta & -\frac{\sqrt{3}}{2} \hat{Q}_\eta & -\frac{\sqrt{3}}{2} \hat{Q}_\zeta & 0 & -\frac{\sqrt{3}}{2} \hat{Q}_\xi \\
    \hat{Q}_\zeta            & 0                                & -\frac{\sqrt{3}}{2} \hat{Q}_\eta & -\frac{\sqrt{3}}{2} \hat{Q}_\xi & 0 \\
 \end{pmatrix}
 \right.
 \nonumber\\
  &+
 \left. 
  \frac{W_E}{2}
 \begin{pmatrix}
 -\{\hat{Q}_E^2\}_u   & \{\hat{Q}_E^2\}_v & 0 & 0 & 0 \\
  \{\hat{Q}_E^2\}_v & \{\hat{Q}_E^2\}_u   & 0 & 0 & 0 \\
  0 & 0 & -\frac{1}{2} \{\hat{Q}_E^2\}_u + \frac{\sqrt{3}}{2} \{\hat{Q}_E^2\}_v & 0 & 0 \\
  0 & 0 & 0 & -\frac{1}{2} \{\hat{Q}_E^2\}_u - \frac{\sqrt{3}}{2} \{\hat{Q}_E^2\}_v & 0 \\
  0 & 0 & 0 & 0 & \{\hat{Q}_E^2\}_u  \\
 \end{pmatrix}
 \right.
 \nonumber\\
 &+ 
 \left.
 \frac{W_E'}{2}
 \begin{pmatrix}
 -\{\hat{Q}_{T_2}^2\}_u   & \{\hat{Q}_{T_2}^2\}_v & 0 & 0 & 0 \\
  \{\hat{Q}_{T_2}^2\}_v & \{\hat{Q}_{T_2}^2\}_u   & 0 & 0 & 0 \\
  0 & 0 & -\frac{1}{2} \{\hat{Q}_{T_2}^2\}_u + \frac{\sqrt{3}}{2} \{\hat{Q}_{T_2}^2\}_v & 0 & 0 \\
  0 & 0 & 0 & -\frac{1}{2} \{\hat{Q}_{T_2}^2\}_u - \frac{\sqrt{3}}{2} \{\hat{Q}_{T_2}^2\}_v & 0 \\
  0 & 0 & 0 & 0 & \{\hat{Q}_{T_2}^2\}_u  \\
 \end{pmatrix}
 + \cdots
 \right]
 \begin{pmatrix}
  \langle u    | \\
  \langle v    | \\
  \langle \xi  | \\
  \langle \eta | \\
  \langle \zeta| \\
 \end{pmatrix}.
 \label{Eq_HJT}
\end{align}
\end{widetext}
Here, 
$\hat{Q}_\gamma$ and $\hat{P}_\gamma$ are the mass-weighted normal coordinates and their conjugate momenta \cite{Wilson1980} for $\Gamma\gamma$ normal mode with frequency $\omega_\Gamma$, 
and $\{\hat{Q}_\Gamma^2\}_{\gamma'}$ are the symmetrized products of the normal coordinates, 
\begin{align}
 \{\hat{Q}_E^2\}_u &= -\frac{1}{\sqrt{2}} \left(\hat{Q}_u^2 - \hat{Q}_v^2\right),
 \nonumber\\
 \{\hat{Q}_E^2\}_v &= \sqrt{2} \hat{Q}_u \hat{Q}_v,
 \nonumber\\
 \{\hat{Q}_{T_2}^2\}_u &= \frac{1}{\sqrt{6}} \left(-\hat{Q}_\xi^2 - \hat{Q}_\eta^2 + 2\hat{Q}_\zeta^2\right),
 \nonumber\\
 \{\hat{Q}_{T_2}^2\}_v &= \frac{1}{\sqrt{2}} \left( \hat{Q}_\xi^2 - \hat{Q}_\eta^2 \right).
\end{align}
$V_\Gamma$ are the linear vibronic coupling parameters, and $W_\Gamma$ are the quadratic vibronic coupling parameters for the $\Gamma$ mode.

The Re $L_3$ edge RIXS measurements of Ba$_2$MgReO$_6$ and their analysis have elucidated several interaction parameters for an embedded $5d^1$ Re ion \cite{Frontini2024}, the
ligand-field splitting parameter $10Dq \approx$ 4.5-5 eV, the spin-orbit coupling parameter $\lambda =0.311$ eV, and the dimensionless vibronic coupling parameter $g_E = V_E/\sqrt{\hslash \omega_E^3} = 1.325$ within the $t_{2g}$ orbitals. 
See Appendix \ref{A_g} for the dimensionless coordinates and vibronic couplings. 
The previous analysis of the experimental data did not provide information on the other vibronic coupling parameters, $V_{T_2}$, $W_E$, and $W_E'$ \cite{Frontini2024}.

\subsection{Adiabatic potential energy surface and vibronic states}
In this section, we discuss the nature of the vibronic eigenstates of the DJT Hamiltonian, $\hat{H}_\text{el} + \hat{H}_\text{DJT}$.
To this end, we consider interactions in descending order of energy scale: the ligand field \eqref{Eq_hLF}, spin-orbit coupling \eqref{Eq_HSO}, and vibronic coupling \eqref{Eq_HJT}. 

The ligand-field splits the $5d$ orbital levels, and spin-orbit coupling further splits them [Fig. \ref{Fig_DP}(b)].
The ligand field lowers the $t_{2g}$ orbital level by $-4Dq$ and raises the $e_g$ orbital level by $6Dq$.
The spin-orbit coupling works within the $t_{2g}$ orbitals and between the $t_{2g}$ and $e_g$ orbitals ($^2T_{2g}$ and $^2E_g$ term states). 
The spin-orbit coupling splits the six $t_{2g}$ spin-orbital states into the $j_\text{eff} = 3/2$ ($\Gamma_8$) and $1/2$ ($\Gamma_7$) multiplet levels with the gap of about $3\lambda/2$, and then slightly shifts the $e_g$ orbital levels and $j_\text{eff}=3/2$ states by about $\lambda^2/10Dq$.  
The resulting electronic energy levels are from the lowest to the higher energy levels, $-4Dq-\frac{1}{2} \lambda$ ($j_\text{eff}=3/2$), $-4Dq+\lambda$ ($j_\text{eff}=1/2$), and $6Dq$ ($^2E_g$). 

The spin-orbit coupling tends to quench the vibronic coupling, while in the present case, the vibronic coupling persists in the four-fold degenerate $j_\text{eff}=3/2$ quartet states \cite{Moffitt1957, Englman1972, Iwahara2024}. 
Let us transform the electronic basis for the vibronic coupling from the $t_{2g}^1$ configurations into the spin-orbit multiplet states $|j_\text{eff}, m_j\rangle$ ($j_\text{eff} = 1/2, 3/2$, and $m_j = -j_\text{eff}, -j_\text{eff}+1, ..., j_\text{eff}$) [Appendix \ref{Sec_t2g_so}]:
\begin{widetext}
\begin{align}
 \hat{V}_\text{JT} &=  
 \left(
  \left|\frac{1}{2},-\frac{1}{2}\right\rangle, 
  \left|\frac{1}{2},\frac{1}{2}\right\rangle, 
  \left|\frac{3}{2},-\frac{3}{2}\right\rangle, 
  \left|\frac{3}{2},-\frac{1}{2}\right\rangle, 
  \left|\frac{3}{2},\frac{1}{2}\right\rangle,
  \left|\frac{3}{2},\frac{3}{2}\right\rangle
 \right)
 \nonumber\\
 &\times 
 \left[
 V_E 
 \begin{pmatrix} 
   0 & 0 & 0 & -\frac{1}{\sqrt{2}} \hat{Q}_v & 0 & \frac{1}{\sqrt{2}} \hat{Q}_u\\
   0 & 0 & -\frac{1}{\sqrt{2}} \hat{Q}_u & 0 & \frac{1}{\sqrt{2}} \hat{Q}_v & 0 \\
   0 & -\frac{1}{\sqrt{2}} \hat{Q}_u & \frac{1}{2} \hat{Q}_u & 0 & \frac{1}{2} \hat{Q}_v & 0 \\
   -\frac{1}{\sqrt{2}} \hat{Q}_v & 0 & 0 & -\frac{1}{2} \hat{Q}_u & 0 & \frac{1}{2} \hat{Q}_v \\
   0 & \frac{1}{\sqrt{2}} \hat{Q}_v & \frac{1}{2} \hat{Q}_v & 0 & -\frac{1}{2} \hat{Q}_u & 0 \\
   \frac{1}{\sqrt{2}} \hat{Q}_u & 0 & 0 & \frac{1}{2} \hat{Q}_v & 0 & \frac{1}{2} \hat{Q}_u \\
 \end{pmatrix} 
 +
 V_{T_2}
 \right.
 \nonumber\\
 &\times
 \left.
 \begin{pmatrix} 
  0 & 0 & \frac{1}{2} \sqrt{\frac{3}{2}} (i \hat{Q}_\xi + \hat{Q}_\eta) & \frac{i}{\sqrt{2}} \hat{Q}_\zeta & \frac{1}{2\sqrt{2}} (-i \hat{Q}_\xi + \hat{Q}_\eta) & 0 \\
  0 & 0 & 0 & \frac{1}{2\sqrt{2}} (i\hat{Q}_\xi + \hat{Q}_\eta) & \frac{i}{\sqrt{2}} \hat{Q}_\zeta & \frac{1}{2} \sqrt{\frac{3}{2}} (-i \hat{Q}_\xi + \hat{Q}_\eta) \\
  \frac{1}{2} \sqrt{\frac{3}{2}} (-i \hat{Q}_\xi + \hat{Q}_\eta) & 0 & 0 & -\frac{1}{2} (i \hat{Q}_\xi + \hat{Q}_\eta) & \frac{i}{2} \hat{Q}_\zeta & 0 \\
  -\frac{i}{\sqrt{2}} \hat{Q}_\zeta & \frac{1}{2\sqrt{2}} (-i \hat{Q}_\xi + \hat{Q}_\eta) & -\frac{1}{2} (-i \hat{Q}_\xi + \hat{Q}_\eta) & 0 & 0 & \frac{i}{2} \hat{Q}_\zeta \\
  \frac{1}{2\sqrt{2}} (i\hat{Q}_\xi + \hat{Q}_\eta) & -\frac{i}{\sqrt{2}} \hat{Q}_\zeta & -\frac{i}{2} \hat{Q}_\zeta & 0 & 0 & \frac{1}{2} (i\hat{Q}_\xi + \hat{Q}_\eta) \\
  0 & \frac{1}{2}\sqrt{\frac{3}{2}} (i\hat{Q}_\xi + \hat{Q}_\eta) & 0 & -\frac{i}{2} \hat{Q}_\zeta & \frac{1}{2} (-i\hat{Q}_\xi + \hat{Q}_\eta) & 0 
 \end{pmatrix} 
 \right]
 \nonumber\\
 &\times
 \left(
  \left\langle\frac{1}{2},-\frac{1}{2}\right|,
  \left\langle\frac{1}{2},\frac{1}{2}\right|,
  \left\langle\frac{3}{2},-\frac{3}{2}\right|,
  \left\langle\frac{3}{2},-\frac{1}{2}\right|,
  \left\langle\frac{3}{2},\frac{1}{2}\right|,
  \left\langle\frac{3}{2},\frac{3}{2}\right|
 \right)^T.
 \label{Eq_HJT_J}
\end{align}
\end{widetext}
Here, superscript $T$ indicates transpose. 
We can readily obtain the quadratic vibronic term by replacing $V$ and $\hat{Q}$ in Eq. \eqref{Eq_HJT_J} with $W/2$ and $\{\hat{Q}^2\}$, respectively. 
The $j_\text{eff} = 3/2$ block of the interaction shows that the spin-orbit coupling reduces the vibronic coupling by half compared with that to the $t_{2g}$ orbitals (\ref{Eq_HJT}), while the vibronic coupling remains nonzero.

The nature of the ground state of the JT model depends on the adiabatic potential energy surface (APES).
To grasp the fundamental information about the APES of the present system, we consider the linear vibronic coupling model within the ground $j_\text{eff}=3/2$ multiplet states. 
For the analysis, we introduce the polar coordinates for the $E_g$ and $T_{2g}$ normal coordinates:
\begin{align}
 (Q_u, Q_v) &= \rho_E (\cos\theta_E, \sin\theta_E), 
\nonumber\\ 
 (Q_\xi, Q_\eta, Q_\zeta) &= \rho_{T_2} (\sin\theta_{T_2} \cos\phi_{T_2}, \sin\theta_{T_2} \cos\phi_{T_2}, \cos\theta_{T_2}). 
 \label{Eq_polar}
\end{align}
Diagonalizing the potential term, we obtain the analytical expressions for the APES.
\begin{align}
 U_\pm &= \frac{\omega_E^2}{2} \rho_E^2 + \frac{\omega_{T_2}^2}{2} \rho_{T_2}^2 \pm 
 \frac{1}{2} \sqrt{V_E^2 \rho_E^2 + V_{T_2}^2 \rho_{T_2}^2}. 
 \label{Eq_Upm}
\end{align}
Each APES represents a set of Kramers doublet levels.
As Eq. (\ref{Eq_Upm}) shows, the APES is independent of the angles in the polar coordinates (\ref{Eq_polar}) within the linear vibronic approximation. 

\begin{figure}[tb]
\begin{tabular}{ll}
(a) & (b) \\
\includegraphics[width=0.47\linewidth, bb = 0 0 297 287]{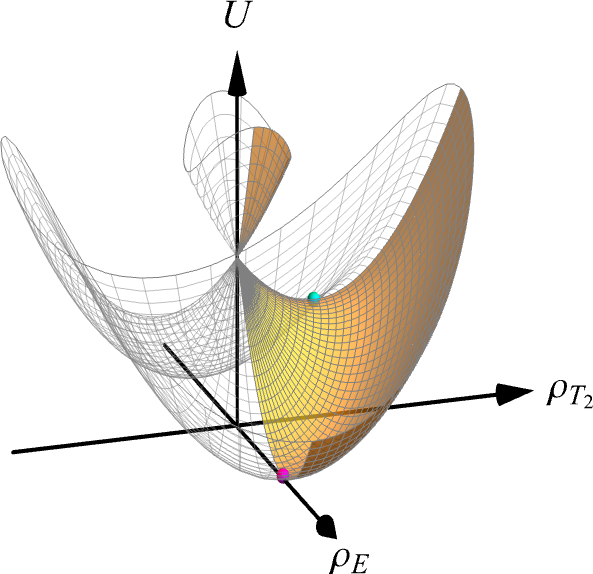}
& 
\includegraphics[width=0.47\linewidth, bb = 0 0 311 277]{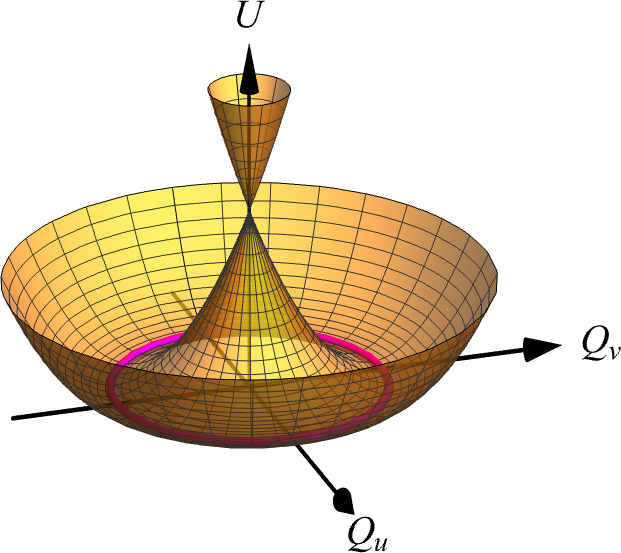}
\end{tabular}
\caption{The adiabatic potential energy surface (APES) for the linear JT model within the $j_\text{eff}=3/2$ multiplet states.
(a) $U_\pm$ with respect to $\rho_E$ and $\rho_{T_2}$ (\ref{Eq_Upm}).
The magenta and cyan points indicate the minimum and the saddle points, respectively.
The plot shows the case for $E_{\text{JT}, E}^{(3/2)} > E_{\text{JT}, T_2}^{(3/2)}$.
(b) The angular dependence of the continuum of minima of $E_g$ type. 
The minima of the lower APES forms a circle (red) around the high-symmetric point. 
In both plots, the origin of the coordinates corresponds to the degeneracy point of the ground electronic state.
}
\label{Fig_APES_LJT}
\end{figure}

Let us analyze the structure of the APES.
Figure \ref{Fig_APES_LJT}(a) is a plot of $U_\pm$. 
In $\rho_E$-$\rho_{T_2}$ coordinate space, the positions and the values of the extrema of $U_\pm$ are 
\begin{widetext}
\begin{align}
 (\rho_E, \rho_{T_2}, U_-) &= 
 \begin{cases}
  \left(\rho_E^\text{JT}, 0, -E_{\text{JT},E}^{(3/2)}\right), 
  & \text{for\;} E_{\text{JT},E}^{(3/2)} > E_{\text{JT},T_2}^{(3/2)}, \\
  \left(\rho_E^\text{JT} \cos x, \rho_{T_2}^\text{JT} \sin x, -E_{\text{JT},E/T_2}^{(3/2)}\right),
  & \text{for\;} E_{\text{JT},E}^{(3/2)} = E_{\text{JT},T_2}^{(3/2)}, \\
  \left(0, \rho_{T_2}^\text{JT}, -E_{\text{JT},T_2}^{(3/2)}\right),
  & \text{for\;} E_{\text{JT},E}^{(3/2)} < E_{\text{JT},T_2}^{(3/2)}.\\
 \end{cases}
 \label{Eq_U_extrema}
\end{align} 
\end{widetext}
Here, $\rho_\Gamma^{\text{JT}}$  ($\Gamma=E, T_{2}$) are the magnitude of the JT deformation, 
\begin{align}
 \rho_\Gamma^\text{JT} &= \frac{V_\Gamma}{2\omega_\Gamma^2},
\end{align}
$E_{\text{JT}, \Gamma}^{(3/2)}$ is the depth of the APES with respect to the ground state in the absence of vibronic coupling, 
\begin{align} 
 E_{\text{JT}, \Gamma}^{(3/2)} = \frac{V_\Gamma^2}{8\omega_\Gamma^2},
 \label{Eq_EJT_j32}
\end{align}
and $0 \le x \le \pi/2$.
For $E_{\text{JT}, E}^{(3/2)} > E_{\text{JT}, T_2}^{(3/2)}$ ($E_{\text{JT}, E}^{(3/2)} < E_{\text{JT}, T_2}^{(3/2)}$), the coordinates at the first and third rows in Eq. (\ref{Eq_U_extrema}) correspond to the global minima and a saddle points (a saddle point and the global minima), respectively.  
Since the APESs of the linear vibronic model do not depend on the angle $\theta_E$ ($\theta_{T_2}$ and $\phi_{T_2}$), the minimum point in Fig. \ref{Fig_APES_LJT}(a) corresponds to a one-dimensional continuum of minima in the $Q_\theta$-$Q_\epsilon$ space [Fig. \ref{Fig_APES_LJT}(b)] (a two-dimensional continuum of minima in the $Q_\xi$-$Q_\eta$-$Q_\zeta$ space).
For $E_{\text{JT}, E}^{(3/2)} = E_{\text{JT}, T_2}^{(3/2)}$, the APES has a continuum of minimum in the $\rho_E$-$\rho_{T_2}$ space too.
In this case, the minima form a four-dimensional surface in the five-dimensional coordinate space. 
See Refs. \cite{Opik1957, Liehr1963, Englman1972, Iwahara2017, Streltsov2020, Streltsov2022}, for further information about the APES.

The previously reported linear vibronic coupling parameters \cite{FioreMosca2024, Zivkovic2024, Soh2024} indicate that the static JT stabilization for the $E_g$ mode is dominant in Ba$_2$MgReO$_6$, i.e., corresponds to the first case in Eq. (\ref{Eq_U_extrema}). 
The APES has a one-dimensional continuum of minima (trough) which is invariant under rotation after $\theta_E$ (\ref{Eq_polar}) as discussed above [Fig. \ref{Fig_APES_LJT}(b)].
The higher-order vibronic and the pseudo JT-type vibronic couplings warp this trough, generating three equivalent minima at the tetragonally compressed structures along one of the three crystallographic axes [Fig. \ref{Fig_DP}(a)].
We will examine the impact of these interactions on the APES in Sec. \ref{Sec_JTE} using the first-principles interaction parameters for a Re ion site in Ba$_2$MgReO$_6$.

The low-energy vibronic eigenstates of $\hat{H}_\text{el} + \hat{H}_\text{DJT}$ delocalize over the three minima of the APES. 
The vibronic wavefunctions basically spread over the warped trough, which leads to the additional stabilization with respect to the one arising from the static JT deformation ($\rho_E > 0$) \cite{Longuet-Higgins1958, Englman1972}.
The vibronic coupling to the $T_{2g}$ modes perturbs the low-energy vibronic states by allowing a slight delocalization of the wave functions over the saddle points of the APES [Fig. \ref{Fig_APES_LJT}(a)].
We will analyze the nature of the vibronic states on Re sites of Ba$_2$MgReO$_6$ in Sec. \ref{Sec_JTE}.

The vibronic states are characterized by the quantum entanglement of the spin-orbit and lattice vibrational degrees of freedom and take the following form \cite{Moffitt1957, Longuet-Higgins1958, Englman1972}:
\begin{align}
 |\Psi_\nu\rangle &= \sum_{i} |\psi_i\rangle \otimes |\chi_{i\nu}\rangle, 
 \label{Eq_vibronic}
\end{align}
where $|\psi_i\rangle$ are the electronic eigenstates, and $|\chi\rangle$ correspond to the lattice part. 
We further determine $|\Psi_\nu\rangle$ in Eq. (\ref{Eq_vibronic}) by solving the corresponding vibronic eigenvalue problem using the first-principles calculated vibronic parameters.

\section{Computational details}
\label{Sec_comput}

\subsection{Electronic states}
\label{Sec_method_CCSD}
We calculated the electronic states of a cluster of Ba$_2$MgReO$_6$ using the equation-of-motion coupled-cluster singles and doubles (EOM-CCSD) method. 
We performed all electronic state calculations using the Q-Chem package (version 6.1) \cite{Qchem}.

\begin{figure}[tb]
\includegraphics[width=0.9\linewidth, bb=0 0 494 475]{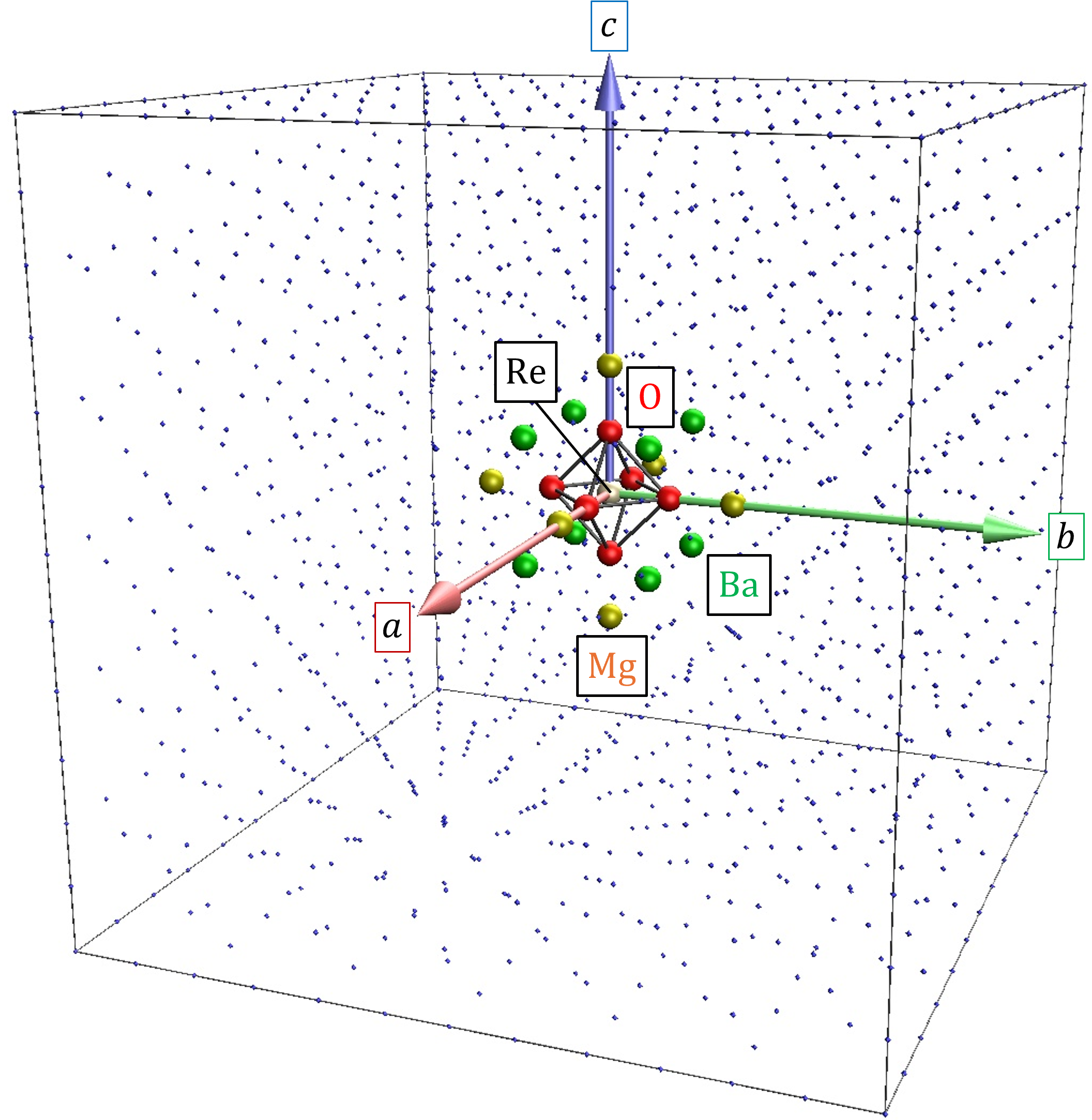}
\caption{
The rhenium cluster embedded in 1420 point charges. 
The cluster consists of ReO$_6$ and the six nearest Mg and eight nearest Ba ions.
The quantum-mechanical cluster is cut out from the crystallographic structure. 
The blue dots represent the point charges. 
}
\label{Fig_cluster}
\end{figure}

The cluster (Fig. \ref{Fig_cluster}) is extracted from the experimental crystal structure \cite{Bramnik2003}. 
The cluster consists of a central ReO$_6$ octahedron, its nearest neighboring diamagnetic ions (eight Ba and six Mg atoms), and point charges representing the surrounding ions. 
We treated the Ba$_8$Mg$_6$ReO$_6$ cluster quantum chemically. 
We included Ba and Mg to accurately describe the Ba-O and Mg-O chemical bondings under JT deformations. 
For these atoms, we employed contracted basis sets as follows: 
For the Re ion, an all-electron type of the exact two-component triple zeta valence and polarization (X2C-TZVP) \cite{Pollak2017}, for the nearest neighbor six oxygen ligand atoms, the correlation-consistent polarized valence triple zeta (cc-pVTZ) basis set \cite{Dunning1989}, and, for the next nearest neighbor Ba and Mg ions, Stevens, Basch, Krauss, Jasien, and Cundari's valence double zeta (SBKJC-VDZ) type with the corresponding effective core potentials for 54 and 10 core electrons, respectively 
\cite{Stevens1984, Stevens1992}. 
These basis sets were obtained from the Basis Set Exchange \cite{Pritchard2019}.
For the Ba$_8$Mg$_6$ReO$_6$
cluster, scalar relativistic effects are 
accounted for with the Hamiltonian of 
spin-free exact two-component theory in its one-electron variant (SFX2C-1e) \cite{Bolvin2017, Ilias2007, Liu2009}.
To account for the contribution from the crystal environment, we replaced all remaining 1420 ions in a supercell composed of $3 \times 3 \times 3$ conventional cells with point charges. 
As charges, we adopted Mulliken charges of Re: $+1.24$, O: $-1.10$, Ba: $+1.84$, and Mg: $+2.11$
from the calculations of Ba$_8$Mg$_6$ReO$_6$ without point charges for the surrounding atoms. Although the Mulliken charges for Re, O, Ba, and Mg change by about 32\%, 9\%, 3\%, and 3\%, respectively, in the presence of the point charges, their impact on the interaction parameters is negligible as discussed in Sec. \ref{Sec_interaction}.  

To describe the $5d^1$ electronic states of the Re cluster, we employed the electron-attachment (EA) approach \cite{Nooijen1995}, in which an extra electron is attached to a $5d^0$ reference state. 
In the calculations, 31 core electrons were frozen in all calculations.
We computed two sets of degenerate term states: ${}^2T_{2g}$ and ${}^2E_g$. 
The energy gap between them corresponds to $10Dq$. 

Using the EOM-CCSD wave functions, we computed the spin-orbit couplings as matrix elements of the Breit-Pauli spin-orbit Hamiltonian, as implemented in the Q-Chem package \cite{Pokhilko2019, Pokhilko2019b, Carreras2020}.
This implementation relies on a mean-field treatment of the two-electron terms and uses the Wigner–Eckart theorem, applied to reduced one-particle density matrices, to derive spin-orbit couplings between all multiplet components \cite{Pokhilko2019}.
We fitted Eq. (\ref{Eq_HSO}) to these matrix elements between the ${}^2T_{2g}$ term states and obtained the spin-orbit coupling parameter $\lambda$.

\subsection{Vibronic coupling parameters}
\label{Sec_method_VW}
We derived the vibronic coupling parameters and frequencies from the EOM-CC APES with respect to the JT-active modes. 
The Cartesian coordinates for the JT deformed structures of the $5d$ metal clusters were calculated by \cite{Wilson1980}
\begin{align}
 \bm{R}_A &= \bm{R}_A^{0} + Q_\alpha \frac{\bm{u}^\alpha_A}{\sqrt{M_A}}, 
\end{align}
where $\bm{R}_A$ are the Cartesian coordinates of atom $A$ in the cluster, $\bm{R}_A^{0}$ the coordinates without the JT deformations corresponding to the experimental structure, $\bm{u}^\alpha$ the normalized eigenvectors of the dynamical matrix, 
and $M_A$ the mass of atom $A$. 
To derive the frequencies and the vibronic couplings to the $E_g$ ($T_{2g}$) modes, we set $Q_u \ne 0$ ($Q_\zeta \ne 0$) and the others to zero. 
These parameters were obtained by fitting the EOM-CCSD ${}^2T_{2g}$ term levels to the potential energy in Eq. (\ref{Eq_HJT}). 

The dimensions of the physical quantities are as follows. 
The dimensions of $\bm{R}$, $M_A$, and $Q_\alpha$ are Bohr radius ($a_0$), the mass of electron ($m_e$), and $a_0 \sqrt{m_e}$.
The dimension of the linear vibronic coupling parameters is $E_h/a_0\sqrt{m_e}$, and of the quadratic vibronic coupling parameters and frequency are $E_h/a_0^2m_e$ and $\sqrt{E_h/a_0^2m_e}$, respectively, where $E_h$ is the Hartree energy. 
For the vibronic coupling parameters, we also use the dimensionless parameters (see Appendix \ref{A_g}).

\subsection{Vibronic states}
\label{Sec_method_DJT}
We calculated the vibronic states by numerically diagonalizing the model vibronic Hamiltonian, $\hat{H}_\text{el} + \hat{H}_\text{DJT}$. 
To this end we expanded $|\chi\rangle$ in Eq. (\ref{Eq_vibronic}) into eigenstates of the harmonic oscillator \cite{Moffitt1957, Longuet-Higgins1958, Muramatsu1978}:
\begin{align}
 |\chi_{i\nu}\rangle &= \sum_{n_u, n_v, n_\xi, n_\eta, n_\zeta=0}^\infty 
 |n_u, n_v, n_\xi, n_\eta, n_\zeta\rangle \chi_{i\bm{n}; \nu},
 \label{Eq_chi}
\end{align}
where $|n_u, n_v, n_\xi, n_\eta, n_\zeta\rangle$ are the eigenstates of the harmonic oscillator part in $\hat{H}_\text{DJT}$, $n_\gamma$ (= 0, 1, 2, ...) are the vibrational quantum numbers for the mode $\gamma$, $\bm{n} = (n_u, n_v, n_\xi, n_\eta, n_\zeta)$, and $\chi_{i\bm{n}; \nu} = \langle \bm{n}|\chi_{i\nu}\rangle$.
With the vibronic basis of $|\psi_i\rangle \otimes |n_u, n_v, n_\xi, n_\eta, n_\zeta\rangle$, we constructed the vibronic Hamiltonian matrix. 

We employed a sufficiently large vibronic basis that can accurately describe the vibronic states and thereby explain the experimental data. 
In our vibronic basis, the vibrational quanta fulfill
\begin{align}
&n_u + n_v + n_\xi + n_\eta + n_\zeta \le 7, 
\nonumber\\
&n_u + n_v \le 5, 
\quad
n_\xi + n_\eta + n_\zeta \le 4. 
\end{align}
We checked the convergence of the energy with respect to the size of the vibrational basis. 
With the DJT model including only the $E_g$ ($T_{2g}$) modes, the ground vibronic level changes by less than $10^{-5}$ eV by increasing the maximum vibrational number to 5 (4). 
The obtained vibronic states are highly accurate to describe the RIXS spectra. 

For the calculations of the vibronic states, we used in-house program code.

\subsection{RIXS spectra}
\label{Sec_method_RIXS}
The obtained vibronic eigenstates were used to simulate the Re $L_3$ edge RIXS spectra. 
In the RIXS process, the embedded Re ion absorbs one photon accompanied by an electronic excitation from $2p_{3/2}$ to $5d$ orbitals, and then emits one photon with the electron transition from a $5d$ orbital to the empty $2p$ orbital.
The initial and final states of the $5d^1$ Re ion are of vibronic type.

We calculate the scattering intensity using the Kramers-Heisenberg formula \cite{Ament2011, Sakurai1967}.
Under the resonant condition, the intensity is, in a good approximation, 
\begin{align}
 I_{fi}(\omega_f) &\propto \left| 
 \sum_n
 \frac{(\bm{\varepsilon}^{(f)}\cdot \langle \Psi_f|\hat{\bm{p}} |n\rangle )(\bm{\varepsilon}^{(i)}\cdot \langle n|\hat{\bm{p}}|\Psi_i\rangle)}{\hslash \omega_i + E_i - E_n + i\frac{\Gamma}{2}} \right|^2
 \nonumber\\
 &\times
 \delta(\hslash \omega_i + E_i -\hslash \omega_f - E_f).
 \label{Eq_I_RIXS}
\end{align}
Here, $\bm{\varepsilon}^{(i)}$ and $\bm{\varepsilon}^{(f)}$ are the polarization vectors of the incident and scattered photons, 
and the corresponding photon energies are $\hslash \omega_i$ and $\hslash \omega_f$; 
the initial ($|\Psi_i\rangle$) and final ($|\Psi_f\rangle$) states are vibronic states of the $5d^1$ octahedron, and $|n\rangle$ are the intermediate states with one core-hole in $2p_{3/2}$, $\Gamma$ being the core-hole lifetime.  
To illustrate the vibronic effect, we compare the spectra arising from the transition between vibronic and pure electronic eigenstates in both the initial and final states.

We approximate 
the intermediate states in Eq. (\ref{Eq_I_RIXS}) as follows. 
The imaginary part ($\Gamma/2$) of Eq. (\ref{Eq_I_RIXS}) for the $2p_{3/2}$ core-hole states of $5d$ metal ions in insulators amounts to as much as 2.75 eV \cite{Clancy2012}. 
The real part of the denominator depends on the type of the $5d^2$ electron configurations.
The $t_{2g}^2$ configurations are close to the resonant condition, and their splitting is much smaller than $\Gamma$, and hence, we ignore the real part and leave only the imaginary part as the denominator of Eq. (\ref{Eq_I_RIXS}). 
The energies of the $t_{2g}^1e_g^1$ configurations are about $10Dq$ higher than those of the $t_{2g}^2$ configurations, and the real part of the denominator is comparable to the imaginary part.  
Within this approximation the cross-section reduces to 
\begin{align}
 I_{fi} &\propto 
 |G_f|^2
 \left| 
 \sum_n
(\bm{\varepsilon}^{(f)}\cdot \langle f|\hat{\bm{p}} |n\rangle )(\bm{\varepsilon}^{(i)}\cdot \langle n|\hat{\bm{p}}|i\rangle)
 \right|^2
 \nonumber\\
 &\times
 \delta(\hslash \omega_i + E_i -\hslash \omega_f - E_f),
 \label{Eq_I_RIXS_FC}
\end{align}
where $G_f$ is the approximate core-hole propagator:
\begin{align}
 |G_f| &= 
 \begin{cases}
  |\Gamma/2|^{-1}, & \text{for the $t_{2g}$ peaks}\\
  |10Dq + i \Gamma/2|^{-1}. & \text{for the $e_{g}$ peaks}
 \end{cases}
\end{align}
We evaluated the transition dipole moments based on the symmetry as in Ref. \cite{Frontini2024}.
We use Eq. (\ref{Eq_I_RIXS_FC}) for the calculations of the $L_3$ edge RIXS spectra. 

Further ingredients for simulating RIXS spectra comply with the experimental conditions \cite{Frontini2024}.
Thus, we used the experimental scattering geometry; 
the wave vectors of the incident ($\bm{k}_i$) and scattered ($\bm{k}_f$) photons are perpendicular to each other, and $\bm{k}_f - \bm{k}_i \parallel [111]$.
The incident and scattered photons are $\pi$-polarized. 
Given that the temperature dependence of the $L_3$ RIXS spectra is minor below 40 K, the RIXS spectra were simulated at $T = 0$ K.
We assumed the Re sites to be completely octahedral because tiny deformations under the quadrupolar ordering do not give significant influence on the shape of the RIXS spectra when averaged over the three domains \cite{Iwahara2025} (for domains in Ba$_2$MgReO$_6$, see Ref. \cite{Muroi2025}).
Finally, we broaden the spectra by convoluting the intensities with Gaussian functions.

\section{Results and discussion}
\label{Sec_result}

\subsection{Electronic and vibronic interaction parameters}
\label{Sec_interaction}

\begin{figure}[tb]
\includegraphics[width=\linewidth, bb = 0 0 504 288]{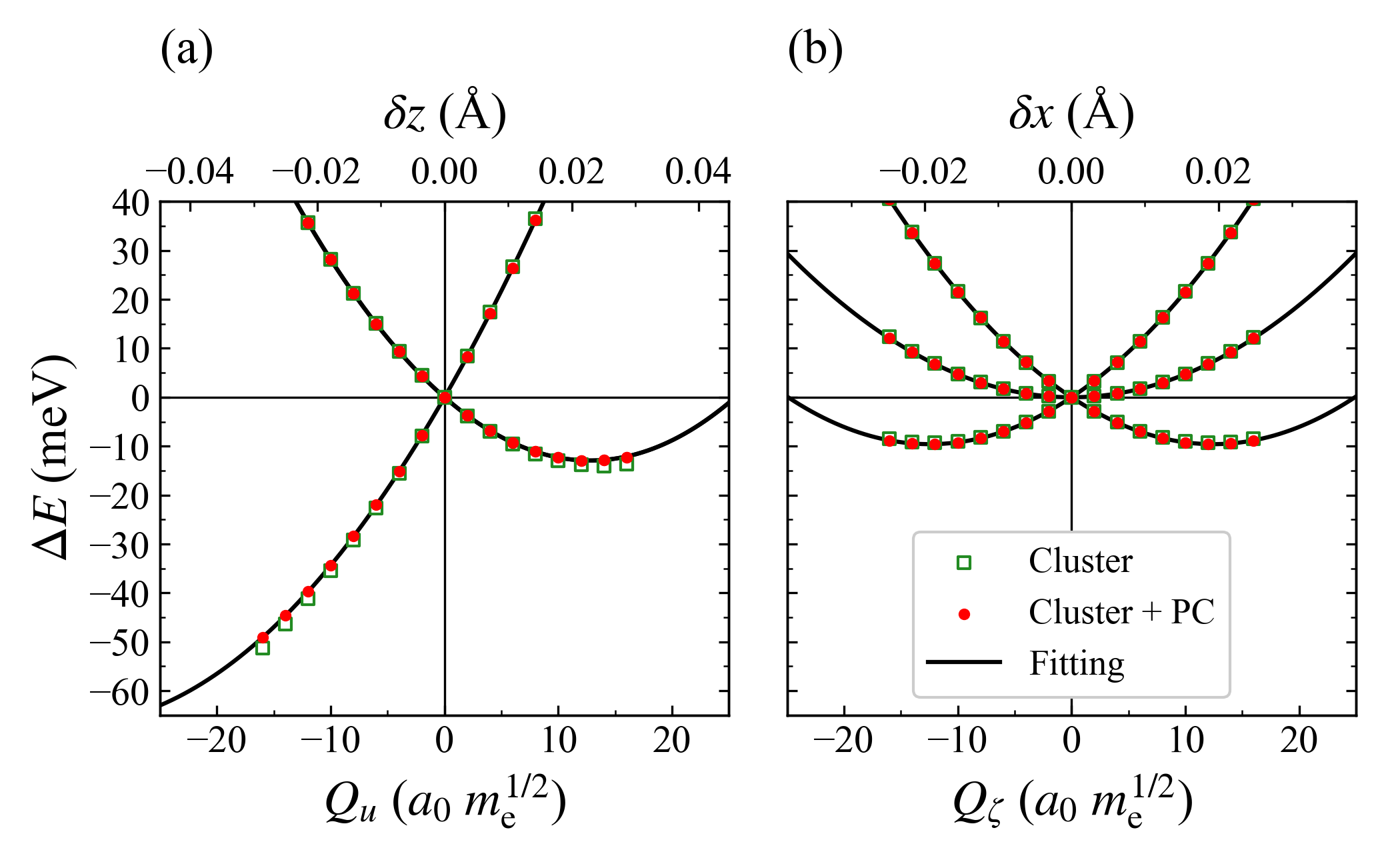}
\caption{The $^2T_{2g}$ adiabatic potential energy surfaces with respect to (a) the $E_gu$ and (b) the $T_{2g}\zeta$ deformations. 
The green squares, the red points, and the black solid lines are, respectively, the $^2T_{2g}$ levels of the Re cluster without surrounding point charges, those of the Re cluster with surrounding point charges (PC), and the adiabatic potential energy surfaces of the model Hamiltonian.
We show the displacements of the oxygen atoms by the JT active modes. 
For $\delta z$ and $\delta x$, see Fig. \ref{Fig_JTmodes}.
}
\label{Fig_APES1}
\end{figure}

\begin{table}[tb]
\caption{Coupling parameters for the $5d^1$ Re cluster. 
The ligand-field and spin-orbit coupling parameters are in eV, and the frequencies and the static and DJT stabilization energies are in meV. 
The vibronic coupling parameters are dimensionless (See Appendix \ref{A_g}).
$E_\text{JT}^{(3/2)}$ corresponds to Eq. \eqref{Eq_EJT_j32}, 
$E_\text{JT}$ is the static JT stabilization energy extracted from the APES, which includes the pseudo JT type vibronic coupling effects, and $E_\text{DJT}$ is the stabilization energy by the DJT effect. 
Figure \ref{Fig_APES2} shows $E_\text{JT}^{(3/2)}$ and $E_\text{JT}$, and Fig. \ref{Fig_level} shows $E_\text{DJT}$.
}
\label{Table_vibronic}
\begin{ruledtabular}
\begin{threeparttable}
\begin{tabular}{cccccccc}
Method & \multicolumn{2}{c}{EOM-CCSD}  & \multicolumn{2}{c}{DFT+HI} & \multicolumn{2}{c}{MRCI} & RIXS \\
       & \multicolumn{2}{c}{This work} & \multicolumn{2}{c}{\cite{FioreMosca2024}} & \multicolumn{2}{c}{\cite{Soh2024}} & \cite{Frontini2024}\\
 \hline                         
                                & \multicolumn{7}{c}{Electronic parameters}\\
$10Dq$                          & \multicolumn{2}{c}{4.88}  & \multicolumn{2}{c}{4.5} & & & \\
$\lambda$                       & \multicolumn{2}{c}{0.321} & \multicolumn{2}{c}{0.29} & & & 0.311 \\
\hline
                                & \multicolumn{7}{c}{Vibronic coupling parameters}\\
$\Gamma$                        & $E_g$ & $T_{2g}$ & $E_g$ & $T_{2g}$ & $E_g$ & $T_{2g}$ & $E_g$ \\
$\omega_\Gamma$                 & $63.1$ & $55.6$ & $75.3$ & $49.8$ & $$ & $$ & $67.3$\\ 
$g_\Gamma$                      & $1.332$ & $0.71$ & $1.20$ & $0.84$ & $$ & $$ & $1.325$\\
$w_\Gamma$                      & $0.23$ & $-0.21$\\
 \hline                         
$E_{\text{JT}, \Gamma}^{(3/2)}$ & 14.2 & 4.4 & 13.6 & 4.4 & 11$^\text{a}$ & 3$^\text{a}$ & 14.7 \\
$E_{\text{JT}, \Gamma}$         & 20.4 & 4.5 &      &     &    &   & 19.1$^\text{b}$ \\
$E_{\text{DJT}}$                & \multicolumn{2}{c}{38.5} & & & & & 31.5$^\text{c}$ \\
\end{tabular}
    \begin{tablenotes}
     \item [a] 1/4 of the reported JT energies for the ${}^2T_{2g}$ term.
     \item [b] Calculated data with $w_E = 0$.
     \item [c] Calculated data with $g_{T_2}=0$ and $w_E = 0$. 
    \end{tablenotes}
\end{threeparttable}
\end{ruledtabular}
\end{table}

We derived the ligand-field parameter from the EOM-CCSD levels and the spin-orbit coupling parameters by mapping the spin-orbit coupling matrix to Eq. \eqref{Eq_HSO}.
Table \ref{Table_vibronic} shows the electronic interaction parameters. 
The energy gap between the ground $j_\text{eff}=3/2$ multiplet level and the $e_g$ level is about $10Dq + \lambda/2 \approx 5.04$ eV, which is in line with the $e_g$ peak position (4.5-5 eV) in the Re $L_3$ edge RIXS spectra \cite{Frontini2024, Zivkovic2024}. 
The $e_g$ term does not show the spin-orbit splitting because the orbital angular momenta are quenched in the term [Sec. 7.3.2 in Ref. \cite{Sugano1970}].
The spin-orbit coupling parameter of 0.321 eV is close to 0.311 eV extracted from the RIXS spectra \cite{Frontini2024}. 
The good agreement between the theoretical calculations and experimental data suggests that the EOM-EA-CC calculation captures a nonnegligible screening upon the addition of a $5d$ electron: 
In terms of the EOM-EA-CC vector, the non-$5d^1$ contributions amount to 6\%.

The frequencies and vibronic coupling parameters were derived from the EOM-CCSD adiabatic potential energy surface.
As described in Sec. \ref{Sec_method_CCSD}, we deformed the Re cluster using the JT active modes and calculated the EOM-CC energy levels at each deformed structure. 
Figure \ref{Fig_APES1} shows the APESs with respect to the $E_g$ and the $T_{2g}$ JT deformations. 
We verified that the present approach gives the frequencies with an accuracy of 5\% by using a diamagnetic analog where Raman scattering data are available \cite{Pasztorova2023} (Appendix \ref{A_BMWO}).

Table \ref{Table_vibronic} lists the calculated frequencies and the vibronic coupling parameters for the embedded Re clusters.
Moreover, the vibronic coupling parameter for the $E_g$ mode ($g_E = $ 1.332) agrees well with the value extracted from the RIXS spectra ($g_E = $ 1.325) \cite{Frontini2024}. 
The calculated vibronic coupling parameter for the $T_{2g}$ mode ($g_{T_2} = 0.71$) shows that the coupling is weaker than that for the $E_g$ mode but is nonnegligible: In terms of the static JT energy [$E_{\text{JT}}^{(3/2)}$, Eq. (\ref{Eq_EJT_j32})], the vibronic coupling to the $T_{2g}$ mode amounts to about 30\% of that to the $E_g$ mode. 
The strengths of the vibronic couplings and $E^{(3/2)}_\text{JT}$'s align with previous calculations using DFT + Hubbard I (DFT+HI) \cite{FioreMosca2024}, and MRCI \cite{Soh2024, Zivkovic2024} approaches.
To evaluate the reliability of the calculated vibronic couplings to the $T_{2g}$ mode, a comparison with experimental data is necessary.

Note that the point charges do not give a large influence on the interaction parameters. 
Without the point charges, we obtained $10Dq=4.80$ eV, $\lambda = 0.322$ eV, 
$\omega_E = 61.5$ meV, $g_E = 1.408$, $w_E = 0.25$, 
$\omega_{T_2} = 56.1$ meV, $g_{T_2} = 0.69$, and $w_{T_2} = -0.21$ 
(see Fig. \ref{Fig_APES1} for the vibronic couplings and frequencies). 
$g_E$ shows the largest deviation of 6\% from the data with the point charges.
The insensitivity of the vibronic coupling to the surrounding point charges supports that vibrations of the nearest oxygen atoms have the dominant effect on the vibronic coupling.

Using the EOM-CC method, we obtained the interaction parameters for $5d^1$ sites in Ba$_2$MgReO$_6$.
The calculated parameters are in agreement with those from the Re $L_3$ edge RIXS spectra \cite{Frontini2024}, which illustrates the reliability of our approach. 
Furthermore, the present approach allows us to extend the model for the embedded Re cluster by including the vibronic coupling to the $T_{2g}$ modes, which have not been extracted from experimental data. 
We examine the validity of the derived vibronic model by simulating the Re $L_3$ edge RIXS spectra in Sec. \ref{Sec_RIXS}.

\subsection{Static and DJT effects}
\label{Sec_JTE}

\begin{figure}[tb]
\includegraphics[width=\linewidth, bb = 0 0 504 288]{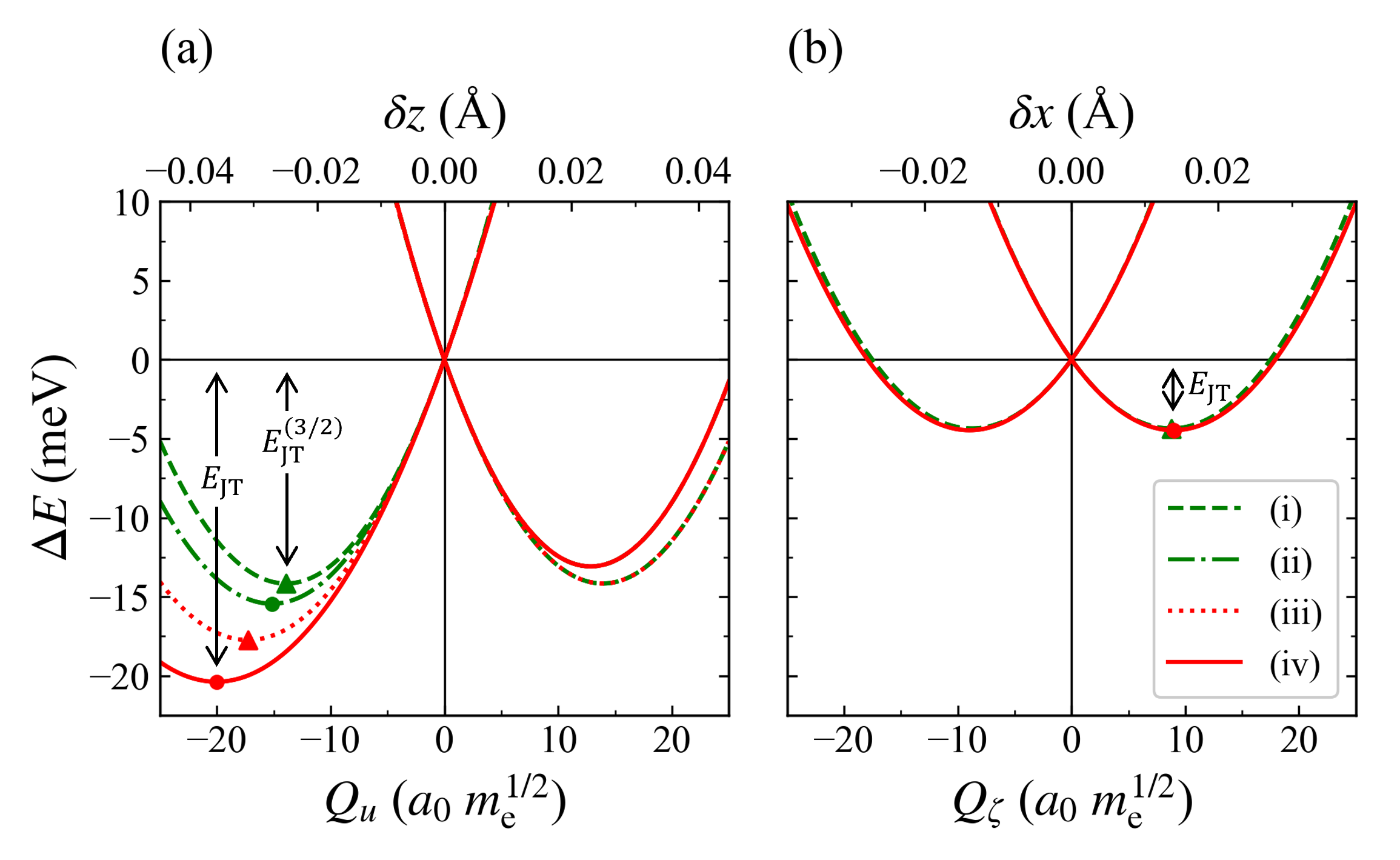}
\caption{The $j_\text{eff}=3/2$ adiabatic potential energy surfaces with respect to (a) the $E_g$ deformations and (b) the $T_{2g}$ deformations.
(i) without quadratic vibronic and pseudo JT (the green dashed line), 
(ii) with quadratic vibronic but without pseudo JT (the green dash-dotted line), 
(iii) without quadratic vibronic but with pseudo JT (the red dotted line), 
(iv) with both the quadratic and pseudo JT couplings (the red solid line). 
For the vibronic models, we used the parameters from the present EOM-CCSD calculations. 
$E_\text{JT}^{(3/2)}$ is the static JT energy of model (i) [Eq. \eqref{Eq_EJT_j32}], and $E_\text{JT}$ is the stabilization energy of model (iv).
}
\label{Fig_APES2}
\end{figure}

\begin{figure}[tb]
\includegraphics[scale=0.45]{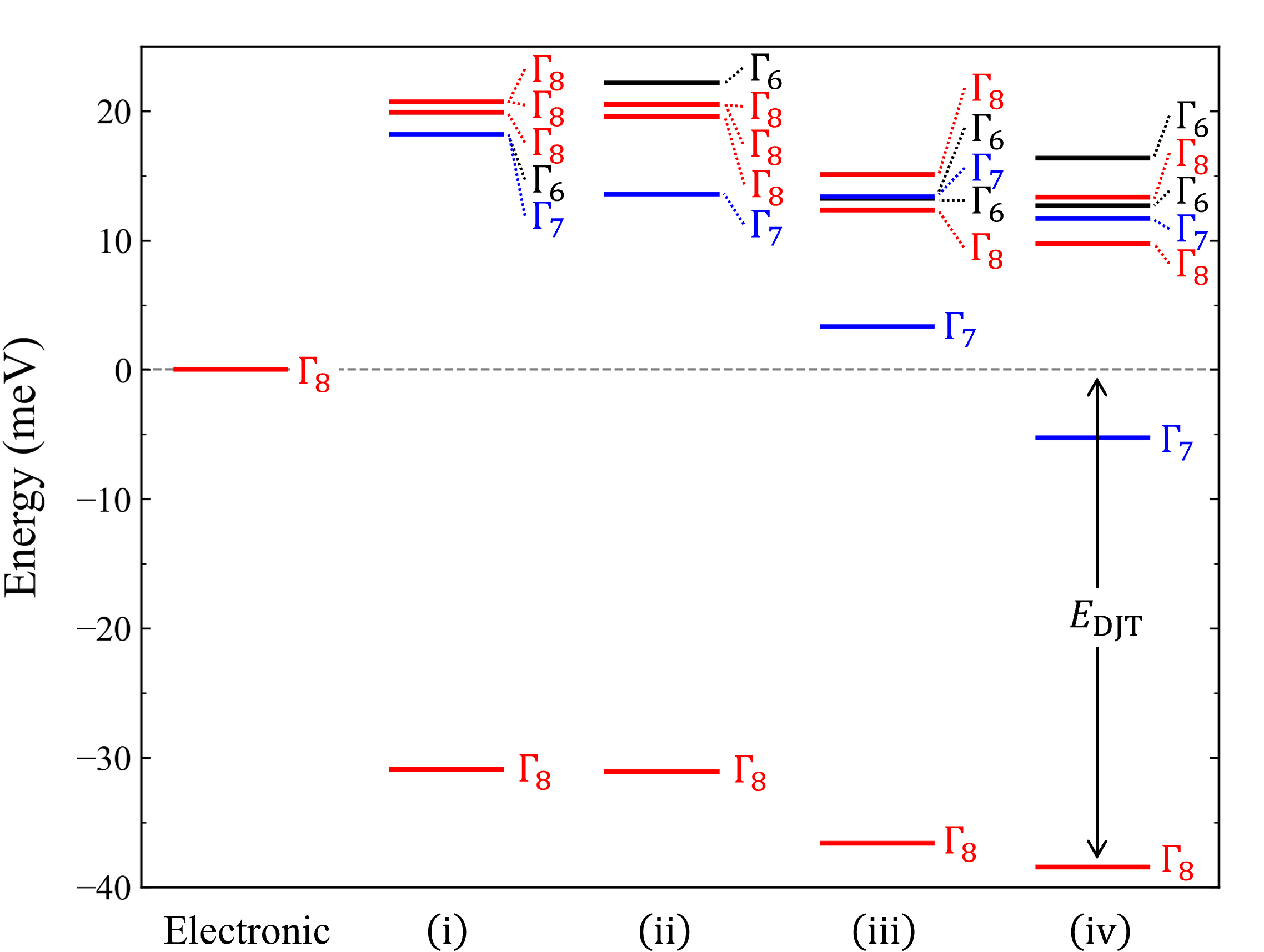}
\caption{The vibronic levels (meV) for DJT models (i)-(iv).
$\Gamma_6$, $\Gamma_7$, and $\Gamma_8$ are the irreducible representations of the vibronic states within the $O_h$ group. 
We set the ground-state energy, in the absence of vibronic coupling, to zero. 
$E_\text{DJT}$ is the stabilization energy by the DJT effect. 
}
\label{Fig_level}
\end{figure}

Now we are in a position to quantitatively analyze the static and the DJT effect on Re sites.
We begin by analyzing the APES of the derived model. 

Figure \ref{Fig_APES2} shows the $j_\text{eff}=3/2$ APESs for the $E_g$ and the $T_{2g}$ deformations. 
To clarify the impact of the quadratic vibronic coupling and pseudo JT coupling between the $j_\text{eff} = 3/2$ and $1/2$ on the JT effect, we analyzed four models:
(i) without pseudo JT and quadratic (the green dashed line), 
(ii) without pseudo JT but with quadratic (the green dash-dotted line), 
(iii) with pseudo JT and without quadratic (the red dotted line), 
and (iv) with both pseudo JT and quadratic (the red solid line) couplings. 
The APES of the linear vibronic model (i) has a continuum of minima in the $Q_u$-$Q_v$ space with depth of $E_{\text{JT}, E}^{(3/2)}$ \eqref{Eq_EJT_j32}. 
The quadratic vibronic and the pseudo JT couplings further stabilize the tetragonally compressed states [Fig. \ref{Fig_JTmodes}(a)] by 1.3 meV and 3.6 meV, respectively. 
Both couplings to the $E_g$ modes enhance the JT deformation in model (iv), for which the total JT stabilization energy ($E_{\text{JT}, E}$) is by 6.2 meV larger than $E_{\text{JT}, E}^{(3/2)}$.
At the minima of the APES within model (iv), the Re-O bond length along one of the crystallographic axes is compressed by 0.036 \AA. 

The APES with respect to the $T_{2g}$ deformations shows that the quadratic and the pseudo JT couplings involving the $T_{2g}$ modes are negligible [Fig. \ref{Fig_JTmodes}(c)]: $E_{\text{JT}, T_2} \approx E_{\text{JT}, T_2}^{(3/2)}$.
Nonlinear vibronic effects become sizable when the linear vibronic coupling is large, which is not the case here. 
Hereafter, we omit the nonlinear vibronic couplings related to the $T_{2g}$ modes.

Quantization of the lattice degrees of freedom in the JT model leads to the stabilization of the spin-orbit-lattice entangled states.
We numerically diagonalized the DJT model ($\hat{H}_\text{el} + \hat{H}_\text{DJT}$) as described in Sec. \ref{Sec_method_DJT} and obtained the energies of these states. 
The ground-state energies of models (i)-(iv) and model (iv) with $V_{T_2}=0$ are stabilized by, respectively, 30.9, 31.1, 36.6, 38.5, and 29.3 meV compared with the ground-state energy of the model without vibronic coupling (Fig. \ref{Fig_level}). 
The most significant contribution to the DJT stabilization energy comes from the linear vibronic coupling to the $E_g$ modes. 
The second largest contribution of $\approx 9.2$ meV comes from the linear vibronic coupling to the $T_{2g}$ modes. 
Then the pseudo JT and the quadratic couplings stabilize the system by about 5.7 meV and 0.2 meV, respectively.

As suggested from the contributions to the DJT stabilization, the vibronic coupling to the $T_{2g}$ modulates the nature of the ground vibronic state.
In the ground vibronic states [Eq. (\ref{Eq_Psi_g_G8})], the vibronic basis states involving $j_\text{eff}=3/2$ and one $T_{2g}$ vibrational excitation amount to as much as 7\%, which is not negligible compared with the contribution of 20 \% from the one $E_g$ vibrational excitation.

Figure \ref{Fig_level} shows how the low-energy vibronic levels evolve with respect to the quadratic vibronic coupling and pseudo JT coupling. 
Some of the low-energy vibronic states are $\Gamma_8 \otimes E$ type, and the others are the hybrid $\Gamma_8 \otimes (E \oplus T_2)$ type. 
The first and second excited ($\Gamma_7$ and $\Gamma_6$) levels of model (i) are of the first type [Eq. (\ref{Eq_Psi_e_G7})].
Introducing the warping of the APES with the quadratic vibronic and the pseudo JT couplings, the $\Gamma_7$ vibronic level lowers \cite{OBrien1964}, approaching the tunneling splitting type excitation in strongly warped APES \cite{Englman1972, Iwahara2024}.
The higher excited vibronic states typically exhibit a hybrid character arising from a mixing of the $E_g$- and $T_{2g}$-type vibrational excitations (the second type).
The first excited $\Gamma_8$ vibronic states above the $\Gamma_7$ level (Fig. \ref{Fig_level}) are mainly of one $T_{2g}$ vibration type, $|j_\text{eff}=3/2 m_j\rangle \otimes |0,0,n_\xi, n_\eta, n_\zeta\rangle$ with $n_\xi + n_\eta + n_\zeta =1$, while these states also have nonnegligible contributions from the $E_g$ vibrationally excited vibronic basis [Eq. (\ref{Eq_Psi_e_G8})].

From the analysis of the JT effect considering both $E_g$ and $T_{2g}$ modes, we showed that the vibronic coupling to the $T_{2g}$ modes gives a nonnegligible contribution to the JT energy (about 20\%) and modulates the nature of the vibronic wave functions.

\begin{figure*}[tb]
\includegraphics[width=\linewidth, bb = 0 0 648 288]{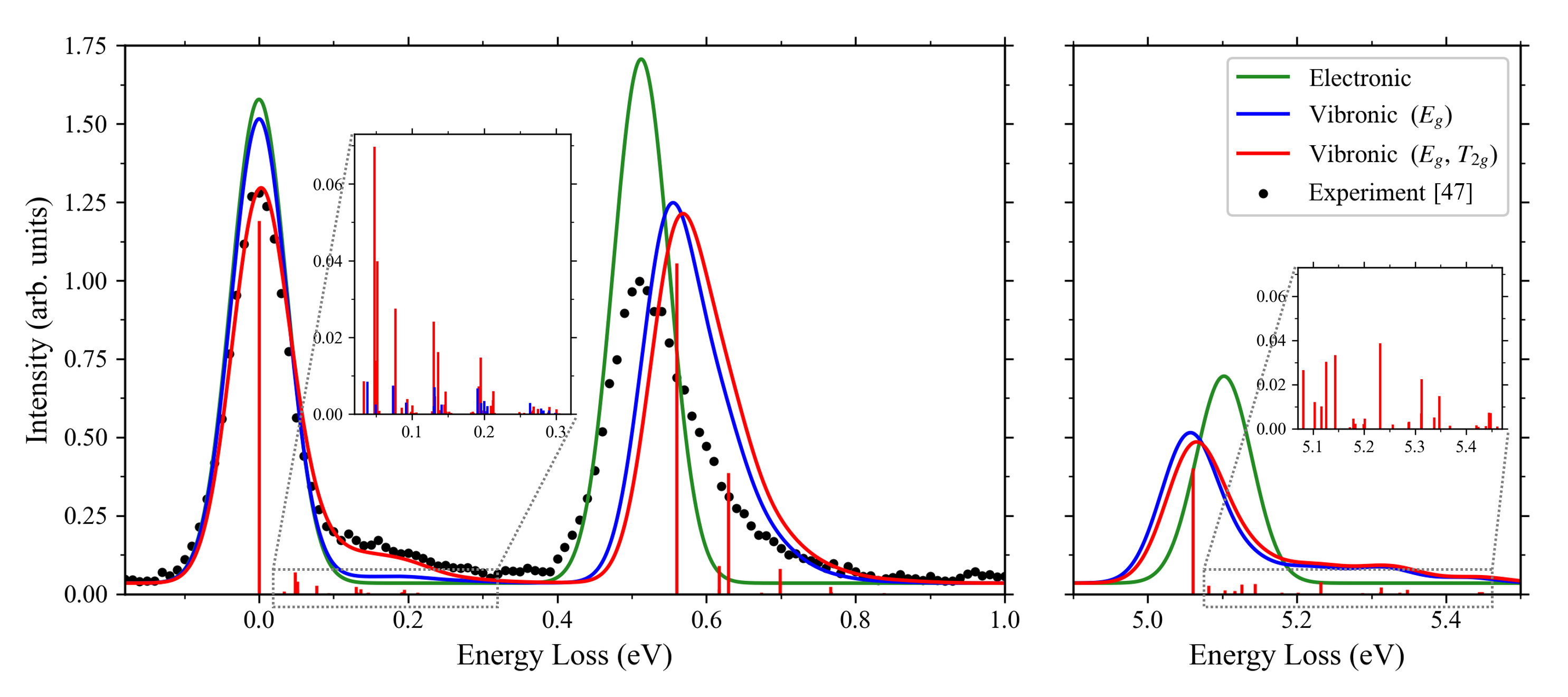}
\caption{Simulated Re $L_3$ edge RIXS spectra.
The green solid lines are the electronic RIXS spectra. 
The blue and red solid lines represent the vibronic RIXS spectra based on models that include only the $E_g$ mode and both the $E_g$ and $T_{2g}$ modes, respectively. 
The black points are the experimental data and taken from Ref. \cite{Frontini2024}.
The insets show the intensities that contribute to the shoulders of the elastic scattering and the $t_{2g}$-$e_g$ transition. 
}
\label{Fig_RIXS}
\end{figure*}

\begin{figure}[tb]
\includegraphics[width=\linewidth, bb = 0 0 504 288]{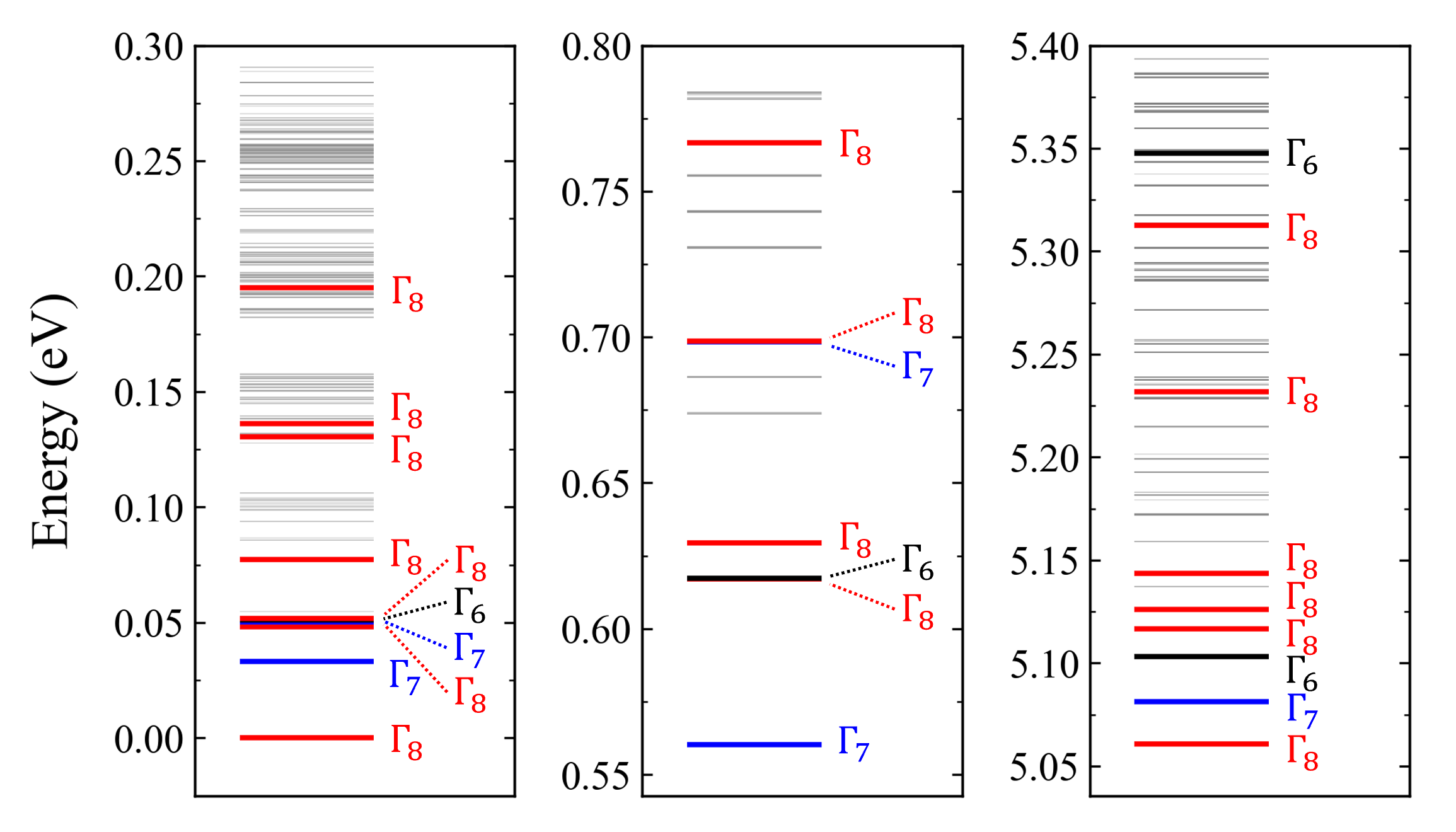}
\caption{The assignment of the RIXS vibronic transitions. The thick lines indicate the vibronic levels relevant to the strong RIXS intensities. 
The thick black, blue, and red lines are the $\Gamma_6$, $\Gamma_7$, and $\Gamma_8$ irreducible representations.
We set the ground level to zero in this figure.
}
\label{Fig_assign}
\end{figure}

\subsection{RIXS spectra}
\label{Sec_RIXS}

With the obtained vibronic states, we simulated the Re $L_3$ edge RIXS spectra (Figure \ref{Fig_RIXS}). 
The electronic RIXS spectra (represented by the green solid lines) have a symmetric peak for each of the electronic levels: the elastic peak at 0 eV, the inelastic peak for $j_\text{eff}=1/2$ at 0.5 eV, and the inelastic peak for $e_g$ (${}^2E_g$) at 5.1 eV.
Turning on the vibronic coupling to the $E_g$ modes, the $j_\text{eff}=1/2$ peak and the $e_g$ peak shift, and their shapes become asymmetric as reported before (the blue solid lines) \cite{Frontini2024, Iwahara2025}.
Then, including vibronic coupling to the $T_{2g}$ modes [model (iv)], the elastic peak shows a clear shoulder below 0.3 eV (the red solid lines).
The low-energy shoulder is in good agreement with the experimentally detected feature in Ref. \cite{Frontini2024}.

The vibronic coupling shifts the peak positions of the $j_\text{eff}=1/2$ and the $e_g$ peaks. 
The DJT effect stabilizes the ground $j_\text{eff}=3/2$ and the $^2E_g$ states. 
The vibronic coupling of the pseudo JT type widens the spin-orbit multiplet levels.
Thus, the $j_\text{eff}=1/2$ peak exhibits a shift towards the higher energy by about 50 meV, while the ${}^2E_g$ peak does towards the lower energy. 

The errors of the calculated peak positions with respect to the experimental ones \cite{Frontini2024} are within 10\%: 
The peaks for the $j_\text{eff} = 1/2$ and ${}^2E_g$ are about 50 meV and 0.5 eV higher than the experimental peaks at 0.50 eV and 4.56 eV, respectively \cite{Frontini2024}. 
The dicrepancy between the experimental and theoretical RIXS spectra could stem from the errors from the experimental data too.
The experimental peak positions in Refs. \cite{Frontini2024} and \cite{Zivkovic2024} for the same compound differ by about 30 meV (6\%). 

Figure \ref{Fig_assign} highlights in color the vibronic levels that contribute strongly to the transitions. 
The asymmetric peaks of the vibronic RIXS spectra come from the multitude of transitions from the ground to the excited vibronic states. 
The mechanism is similar to Franck-Condon's for shifted oscillators: a multitude of excitations from the ground to the excited vibronic levels give rise to the asymmetric peaks. 
One can find a detailed description of the asymmetric $j_\text{eff}=1/2$ peak in Refs. \cite{Frontini2024, Iwahara2025}.
The broad shoulder of the elastic peak appears due to the transitions from the ground vibronic levels to the excited vibronic levels that mainly arise from the $T_{2g}$ vibrational excitations, Eq. \eqref{Eq_Psi_e_G8}. 

With the present model, we have reproduced all the essential features of the Re $L_3$ edge RIXS spectra. 
In particular, the shoulder peak arises from the vibronic coupling to the $T_{2g}$ modes, which was missing in earlier models.

\section{Conclusion}
In this work, we investigated the DJT effect for $5d^1$ Re sites in Ba$_2$MgReO$_6$ by using the EOM-CC method, and simulated the Re $L_3$ edge RIXS spectra. 
The calculated ligand-field splitting, spin-orbit coupling, and vibronic couplings are in close agreement with the reported experimental data \cite{Frontini2024}: The discrepancies are within 5\%, which illustrates the reliability of the adopted CC approach. 
From the simulated RIXS spectra, the shoulder on the elastic peak is induced by the vibronic coupling to the $T_{2g}$ vibrations, which was not assigned in the previous analysis due to the neglect of $T_{2g}$ vibrational degrees of freedom. 
The present results illustrate that, for a unified quantitative description of the multipolar features of the family of $5d^1$ double perovskites, the DJT effect is crucial. 

The present work also demonstrates the usefulness of the EOM-CC method in elucidating the local vibronic model and analyzing the RIXS spectra of the $5d$ compound. 
The EOM-CC-based approach will apply to various transition-metal compounds, e.g., those with low symmetry and more $d$ electrons. 
The present protocol using the EOM-CC method opens an avenue to accurate theoretical predictions of the complex electronic states of various correlated quantum materials and their spectroscopic signatures.

\begin{acknowledgments}
T.M. and N.I. are grateful to the Division of Quantum and Physical Chemistry at KU Leuven for hospitality. 
The work at Chiba was partly supported by 
Grant-in-Aid for Scientific Research (Grant No. 22K03507) from the Japan Society for the Promotion of Science, 
the Nippon Sheet Glass Foundation for Materials Science and Engineering, 
and the JST CREST project led by T. Omatsu (No. JPMJCR1903).
\end{acknowledgments}

\section*{Data Availability}
The data that support the findings of this article are openly available \cite{data}.

\appendix
\section{Dimensionless vibronic quantities}
\label{A_g}
Dimensionless vibronic coupling parameters are convenient for theoretical descriptions of the DJT effect. 
We transform the mass-weighted normal coordinates and the conjugate momenta \cite{Wilson1980} into the dimensionless coordinates and momenta, respectively, by the following relations:
\begin{align}
 \hat{Q}_{\gamma} = \sqrt{\frac{\hslash}{\omega_{\Gamma}}} \hat{q}_{\gamma},
 \quad 
 \hat{P}_{\gamma} = \sqrt{\hslash\omega_{\Gamma}} \hat{p}_{\gamma}.
\end{align}
The dimensionless coordinates and momenta are, in terms of the vibrational creation and annihilation operators ($\hat{b}_\gamma^\dagger$ and $\hat{b}_\gamma$), 
\begin{align}
 \hat{q}_\gamma = \frac{1}{\sqrt{2}} \left( \hat{b}_\gamma^\dagger + \hat{b}_\gamma \right),
 \quad
 \hat{p}_\gamma = \frac{i}{\sqrt{2}} \left( \hat{b}_\gamma^\dagger - \hat{b}_\gamma \right).
\end{align}

Substituting them into the vibronic Hamiltonian, the Hamiltonian reduces to the form of the products of the frequency ($\hslash \omega$) and dimensionless operator. 
The latter contains dimensionless vibronic coupling parameters defined by 
\begin{align}
 g_{\Gamma} = \frac{V_{\Gamma}}{\sqrt{\hslash \omega_\Gamma^3}}, \quad 
 w_{E} = \frac{W_{E}}{\omega_E^2}, 
 \quad 
 w'_{E} = \frac{W'_{E}}{\omega_{T_2}^2}.
\end{align}
We obtain a dimensionless form of the vibornic Hamiltonian by replacing $V$ ($W$) and $\hat{Q}$ with $\hslash \omega g$ ($\hslash \omega w$) and $\hat{q}$, respectively.

\section{$t_{2g}^1$ configurations and spin-orbit multiplet states}
\label{Sec_t2g_so}
The relation between the $t_{2g}^1$ electronic configurations and the spin-orbit multiplet states is 
\begin{align}
 |j_\text{eff}, m_j\rangle &= \sum_{\gamma\sigma} |\gamma \sigma\rangle (\Gamma_5\gamma, \Gamma_6 \sigma|j_\text{eff}, m_j),
\end{align}
where Clebsch-Gordan coefficients for the $O_h$ group, $(\Gamma_5\gamma, \Gamma_6 \sigma|j_\text{eff}, m_j)$ \cite{Koster1963}, were used; $j_\text{eff}=1/2$ and $3/2$ are, respectively, $\Gamma_7$ and $\Gamma_8$ representations. 
The explicit forms of the relation are as follows:
\begin{align}
&
 \left(
  \left|\frac{1}{2},-\frac{1}{2}\right\rangle, 
  \left|\frac{1}{2},\frac{1}{2}\right\rangle, 
  \left|\frac{3}{2},-\frac{3}{2}\right\rangle, 
  \left|\frac{3}{2},-\frac{1}{2}\right\rangle, 
  \left|\frac{3}{2},\frac{1}{2}\right\rangle,
  \left|\frac{3}{2},\frac{3}{2}\right\rangle
 \right)
 \nonumber\\
 &= 
 \left(
  \left|\xi,   \downarrow\right\rangle, 
  \left|\xi,   \uparrow\right\rangle, 
  \left|\eta,  \downarrow\right\rangle, 
  \left|\eta,  \uparrow\right\rangle, 
  \left|\zeta, \downarrow\right\rangle,
  \left|\zeta, \uparrow\right\rangle
 \right)
 \bm{U},
\\
 \bm{U} &=
  \begin{pmatrix}
    0& -\frac{i}{\sqrt{3}} & 0 & -\frac{1}{\sqrt{3}} & \frac{i}{\sqrt{3}} & 0 \\
    -\frac{i}{\sqrt{3}} & 0 & \frac{1}{\sqrt{3}} & 0 & 0 & -\frac{i}{\sqrt{3}} \\
    -\frac{i}{\sqrt{6}} & 0 & \frac{1}{\sqrt{6}} & 0 & 0 & i \sqrt{\frac{2}{3}} \\
    0 & \frac{i}{\sqrt{2}} & 0 & -\frac{1}{\sqrt{2}} & 0 & 0 \\
    -\frac{i}{\sqrt{2}} & 0 & -\frac{1}{\sqrt{2}} & 0 & 0 & 0\\
    0 & \frac{i}{\sqrt{6}} & 0 & \frac{1}{\sqrt{6}} & i\sqrt{\frac{2}{3}} & 0 \\
  \end{pmatrix}.
\end{align}

\section{Frequencies of Ba$_2$MgWO$_6$}
\label{A_BMWO}

\begin{figure}[tb]
\includegraphics[width=\linewidth, bb = 0 0 504 288]{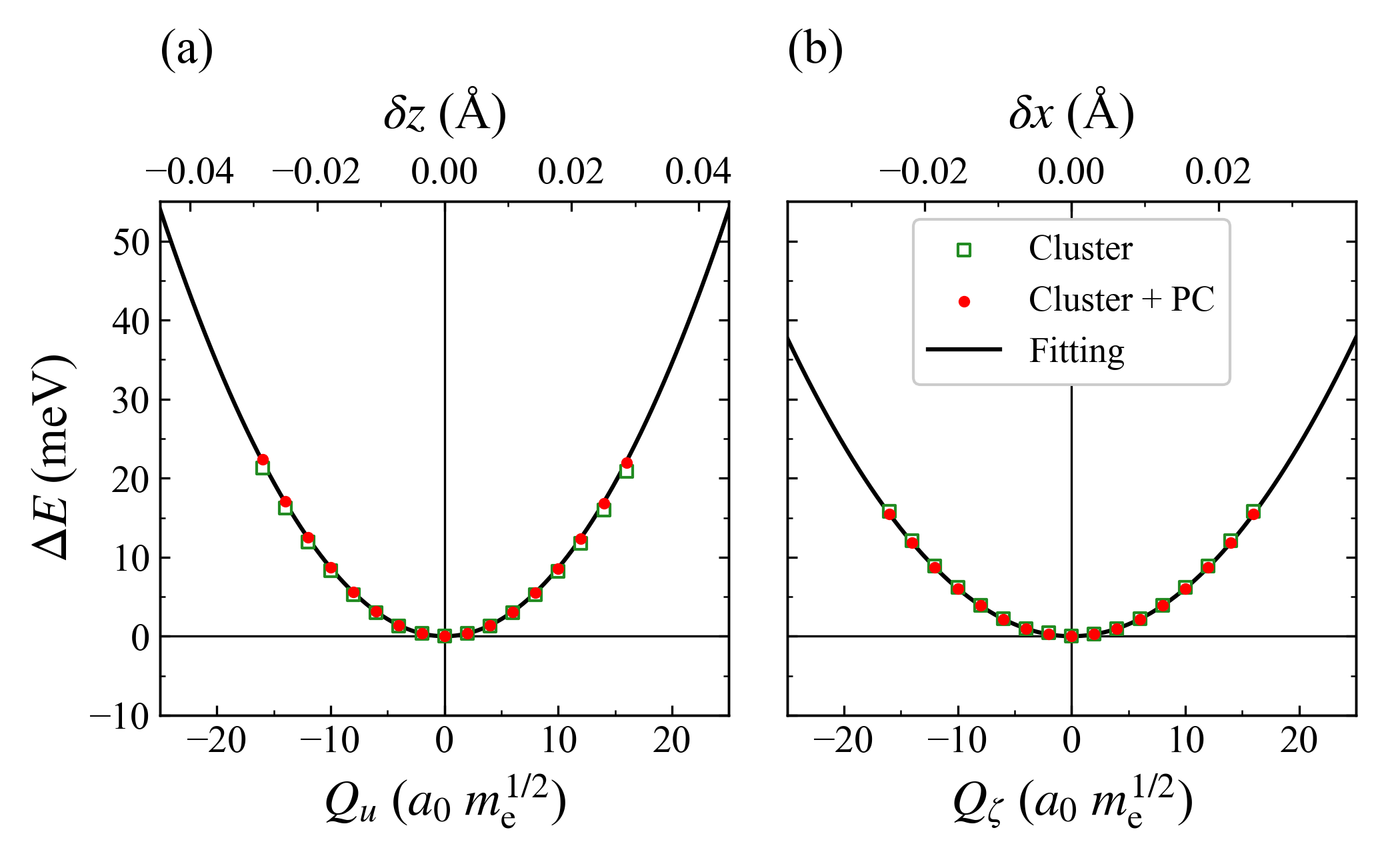}
\caption{The total energies of the $5d^0$ W cluster with respect to (a) the $E_g$ and (b) the $T_{2g}$ deformations. 
The green squares, the red points, and the black solid lines are, respectively, the CCSD data for the W cluster without surrounding point charges, the CCSD data for the W cluster with surrounding point charges (PC), and the fit harmonic potential.}
\label{Fig_d0}
\end{figure}

We examined the accuracy of the frequencies obtained by the CCSD method by calculating the frequencies of a $5d^0$ W cluster of nonmagnetic Ba$_2$MgWO$_6$. 
We used the latter because its experimental frequencies are available \cite{Pasztorova2023}, unlike those of Ba$_2$MgReO$_6$. 
We constructed the W-based cluster (Fig. \ref{Fig_cluster}) from the experimental structure of the crystal \cite{Pasztorova2023}.
The point charges were W: $+2.37$, O: $-1.25$, Ba: $+1.82$, and Mg: $+2.10$. 
The HF and CCSD calculations were performed using the same basis sets and computational approximations as for the Re cluster. 

Figure \ref{Fig_d0} shows the APESs of the $5d^0$ W cluster of Ba$_2$MgWO$_6$. 
The obtained frequencies are 68.6 meV and 57.3 meV for the $E_g$ and $T_{2g}$ modes, respectively, 
which are in good agreement with the Raman scattering data of 67.3 meV and 54.4 meV \cite{Pasztorova2023}.

\section{Vibronic states}
\label{A_wf}
Here, we give the dominant contributions for several vibronic wave functions of the DJT model (iii). 
The ground vibronic state with the largest projection is 
\begin{align}
|\Psi_{\Gamma_8,-\frac{3}{2}}\rangle 
&= 
 0.82\left|-\frac{3}{2}\right\rangle |0\rangle 
 -0.45 
 \left|\Phi_{\Gamma_8,-\frac{3}{2}}^{\Gamma_8 \otimes E}\right\rangle
 \nonumber\\
 &-0.26
 \left(
  \frac{1}{\sqrt{5}} \left|\Phi_{\Gamma_8^{(1)},-\frac{3}{2}}^{\Gamma_8 \otimes T_2}\right\rangle 
- \frac{2}{\sqrt{5}} \left|\Phi_{\Gamma_8^{(2)},-\frac{3}{2}}^{\Gamma_8 \otimes T_2}\right\rangle 
\right)
 + \cdots,
\label{Eq_Psi_g_G8}
\end{align}
where $|m_j\rangle$ stand for $|j_\text{eff}=3/2, m_j\rangle$ multiplet states, 
and the symmetrized $\Gamma_8$ vibronic basis states with single vibrational excitation are 
\begin{align}
 \left|\Phi_{\Gamma_8,-\frac{3}{2}}^{\Gamma_8\otimes E}\right\rangle &= 
 \frac{1}{\sqrt{2}} 
 \left( 
  \left|{-\frac{3}{2}}\right\rangle |1_u\rangle 
  +
  \left|{\frac{1}{2}} \right\rangle |1_v\rangle 
 \right),
 \\
 \left|\Phi_{\Gamma_8^{(1)},-\frac{3}{2}}^{\Gamma_8 \otimes T_2}\right\rangle &= 
 i\frac{2}{\sqrt{15}}  \left|-\frac{1}{2}  \right\rangle|1_\xi\rangle 
 + \frac{i}{\sqrt{5}}  \left|{\frac{3}{2}} \right\rangle |1_\xi\rangle 
 - \frac{2}{\sqrt{15}} 
 \nonumber\\
&\times
 \left|{-\frac{1}{2}}\right\rangle |1_\eta\rangle 
  + \frac{1}{\sqrt{5}}  \left|{\frac{3}{2}} \right\rangle |1_\eta\rangle 
  + \frac{i}{\sqrt{15}} \left|{\frac{1}{2}} \right\rangle |1_\zeta\rangle,
 \label{Eq_e_G8_T2_A}
 \\
 \left|\Phi_{\Gamma_8^{(2)},-\frac{3}{2}}^{\Gamma_8 \otimes T_2}\right\rangle &= 
 -i\sqrt{\frac{3}{20}}  \left|{-\frac{1}{2}}\right\rangle|1_\xi\rangle 
 + \frac{i}{\sqrt{20}}  \left|{\frac{3}{2}} \right\rangle |1_\xi\rangle 
 + \sqrt{\frac{3}{20}}  
 \nonumber\\
&\times
 \left|{-\frac{1}{2}}\right\rangle |1_\eta\rangle 
 + \frac{1}{\sqrt{20}}  \left|{\frac{3}{2}} \right\rangle |1_\eta\rangle 
 + i \sqrt{\frac{3}{5}} \left|{\frac{1}{2}} \right\rangle |1_\zeta\rangle.
 \label{Eq_e_G8_T2_B}
\end{align}
$|1_\gamma\rangle$ indicates the vibrational states with one vibrational excitation of mode $\gamma$.
We constructed these symmetrized vibronic basis sets by using the Clebsch-Gordan coefficients \cite{Koster1963}.

The lowest $\Gamma_7$ vibronic states originate from the $E_g$ vibrationally excited configurations. 
\begin{align}
 |\Psi_{\Gamma_7,+\frac{1}{2}}\rangle &= 
 0.79 
 \left[
 \frac{1}{\sqrt{2}} \left(\left|{\frac{3}{2}}\right\rangle |1_u\rangle - \left|{-\frac{1}{2}}\right\rangle |1_v\rangle \right) 
 \right]
 + \cdots.
\label{Eq_Psi_e_G7}
\end{align}

The first excited $\Gamma_8$ vibronic states mainly come from the $T_{2g}$ vibrationally excited states, while containing nonnegligible contributions from the $E_g$ vibrationally excited states:
\begin{align}
 |\Psi'_{\Gamma_8,-\frac{3}{2}}\rangle 
 &= 0.42 \left|\Psi_{\Gamma_8,-\frac{3}{2}}^{\Gamma_8 \otimes E}\right\rangle 
 - 0.05  \left|\Psi_{\Gamma_8^{(1)},-\frac{3}{2}}^{\Gamma_8 \otimes T_2}\right\rangle 
 \nonumber\\
 &+ 0.63  \left|\Psi_{\Gamma_8^{(2)},-\frac{3}{2}}^{\Gamma_8 \otimes T_2}\right\rangle 
 + \cdots.
\label{Eq_Psi_e_G8}
\end{align}


%
\end{document}